\documentclass[10pt,
floatfix, 
aps,
pra, 
amsmath,
twocolumn,
superscriptaddress,
]{revtex4-2}

\pdfoutput=1

\usepackage{graphicx}
\usepackage{amsmath}
\usepackage{amssymb}
\usepackage{bbm}
\usepackage{soul}
\usepackage{xcolor}
\usepackage{bbold}
\usepackage{physics}
\usepackage{mathtools}
\usepackage{booktabs}
\usepackage{ulem}
\usepackage[unicode=true, colorlinks=true]{hyperref}

\hypersetup{
    linkcolor=blue,
    citecolor=blue,
    urlcolor=blue
}

\begin{document}

\title{Quasiparticle-induced transitions in a fluxonium qubit}

\author{Maksim Litskevich}
\thanks{These authors contributed equally to this work.}
\affiliation{Department of Physics, Syracuse University, Syracuse, NY 13244, USA}
\affiliation{Institute for Quantum and Information Sciences,
Syracuse University, Syracuse, NY 13244, USA}

\author{Kesavan Manivannan}
\thanks{These authors contributed equally to this work.}
\affiliation{Department of Physics, Syracuse University, Syracuse, NY 13244, USA}

\author{Benjamin Byrd}
\affiliation{Department of Physics, Syracuse University, Syracuse, NY 13244, USA}

\author{Pavel D. Kurilovich}
\affiliation{Department of Physics, Harvard University, Cambridge, Massachusetts 02138, USA}

\author{Vladislav D. Kurilovich}
\affiliation{Google Quantum AI, Santa Barbara, CA 93117, USA}

\author{Gianluigi Catelani}
\affiliation{Institute for Theoretical Nanoelectronics (PGI-2), Forschungszentrum Jülich, 52428 Jülich, Germany}
\affiliation{Quantum Research Center, Technology Innovation Institute, Abu Dhabi 9639, United Arab Emirates}

\author{Ivan V. Pechenezhskiy}
\email{ivpechen@syr.edu}
\affiliation{Department of Physics, Syracuse University, Syracuse, NY 13244, USA}
\affiliation{Institute for Quantum and Information Sciences,
Syracuse University, Syracuse, NY 13244, USA}

\date{\today}

\begin{abstract}
Quasiparticles are a prominent decoherence source in superconducting qubits, but their effects are notoriously difficult to isolate in fluxonium. Unlike the transmon, fluxonium is insensitive to offset charge, precluding charge-parity detection of quasiparticle tunneling. We address this challenge by measuring the excitation and de-excitation rates in a fluxonium qubit under controlled on-chip quasiparticle injection and isolating the quasiparticle-induced contribution by subtracting the background transition rates. We show that to accurately model the external magnetic flux dependence of the quasiparticle-induced transition rates, it is necessary to account for the superconducting gap asymmetry across the Josephson junctions. A comparison between theory and experiment constrains the relative quasiparticle densities in the small junction and the junction array and helps explain previously reported discrepancies between the bounds on the quasiparticle densities inferred for these two circuit elements.

\end{abstract}

\maketitle

\section{Introduction}
\label{sec:introduction}

Superconducting qubits are among the leading platforms for quantum computing, with individual physical qubit error rates below the quantum error correction threshold~\cite{Acharya2025}. However, quantum error correction schemes are susceptible to correlated errors induced by quasiparticle bursts caused by ionizing radiation impacts~\cite{Wilen2021, McEwen2022, McEwen2024, Harrington2025, KurilovichV2026} or mechanical stress relief~\cite{AnthonyPetersen2024, Yelton2025}. In the aftermath of such events, a significant fraction of the deposited energy is converted into Cooper-pair-breaking (CP-breaking) phonons~\cite{Martinis2021}. These phonons, in turn, create Bogoliubov quasiparticles (QPs) upon hitting the superconducting (SC) films that form the electrodes of the Josephson junctions, thereby causing decoherence errors~\cite{Lutchyn2005, Martinis2009, Catelani2011, Catelani2012}. Extensive studies of QP-induced decoherence in transmon qubits led to the development of mitigation strategies, including the use of normal-metal reservoirs on the backside of the chip for phonon down-conversion~\cite{Iaia2022} and engineered SC energy gap profiles to suppress QP tunneling~\cite{Marchegiani2022, McEwen2024, KurilovichV2026, Pinckney2026}.

Meanwhile, the fluxonium qubit has been explored as an alternative to the transmon qubit due to its long coherence times and large anharmonicity~\cite{Nguyen2019, Nguyen2022, Somoroff2023, Ardati2024, Wang2025}. In a fluxonium qubit, a small Josephson junction is shunted with a large inductor, typically implemented with an array of larger Josephson junctions~\cite{Manucharyan2009}. In such a circuit, QP-induced qubit decoherence arises from QP tunneling in the small Josephson junction and the junction array~\cite{Pop2014, Vool2014}.

Several previous studies inferred \textit{bounds} on the reduced QP density $x_\mathrm{qp} = n_\mathrm{qp} / n_\mathrm{CP}$, where $n_\mathrm{qp}$ and $n_\mathrm{CP}$ are the QP and CP volume densities~\cite{Catelani2011}, by measuring fluxonium energy-relaxation times~$T_1$ and attributing all or some fraction of the loss to QPs~\cite{Pop2014, Vool2014, Grunhaupt2019, Nguyen2019, Somoroff2023, Atanasova2025, Watanabe2025, Ateshian2025, Azar2026Exp, Larson2026, Zhuang2026}. Somewhat unexpectedly, these studies have consistently inferred a higher upper bound on the QP density in the small junction, $x_\mathrm{qp}^\mathrm{small}$, than in the array, $x_\mathrm{qp}^\mathrm{array}$, with $x_\mathrm{qp}^\mathrm{small}$ often exceeding $x_\mathrm{qp}^\mathrm{array}$ by at least an order of magnitude, as summarized in Appendix~\ref{app:xqp_literature} and Table~\ref{tab:qp_density_summary}. This includes Ref.~\cite{Pop2014}, where a bound on $x_\mathrm{qp}^\mathrm{array}$ can be obtained from the bound on inductive losses reported in that work.

Crucially, the bounds on $x_\mathrm{qp}$ in fluxonia are typically estimated using expressions for QP-induced energy relaxation derived under the assumptions $\delta E_\mathrm{qp} \ll \hbar\omega_{01}$ and $\Delta_L = \Delta_H$~\cite{Catelani2011, Glazman2021, Pop2014}. Here, $\omega_{01}$ is the angular transition frequency between the ground $\ket{0}$ and excited $\ket{1}$ qubit states, $\delta E_\mathrm{qp}$ is the characteristic energy width of the QP distribution, $\Delta_L$ and $\Delta_H$ are SC energy gaps in the low- and high-gap Josephson junction leads, and $h = 2\pi\hbar$ is the Planck constant. The first approximation, $\delta E_\mathrm{qp} \ll \hbar\omega_{01}$, is generally valid near the integer flux quantum (IFQ) bias point, where fluxonium operates in a transmon-like regime with a relatively high $\omega_{01}$, but it can break down near the half-integer flux quantum (HFQ) point, where $\hbar\omega_{01}$ reaches its minimum and can be comparable to or even smaller than $\delta E_\mathrm{qp}$. The second approximation, $\Delta_L = \Delta_H$, should not be assumed for typical qubit-junction fabrication processes, where the two SC electrodes often have different thicknesses. This thickness difference results in an SC gap asymmetry between the junction leads $\delta\Delta = \Delta_H - \Delta_L$~\cite{Marchegiani2022, Connolly2024}. The gap asymmetry can significantly exceed $\hbar\omega_{01}$ near the HFQ, where fluxonium is commonly operated. It therefore remains an open question whether the reported disparity between bounds on the QP densities in the small junction and the junction array reflects a genuine difference in QP densities in these circuit elements or arises from the simplifying assumptions used to derive those bounds.

In this work, we investigate QP-induced transition rates in fluxonium qubits under controlled on-chip QP injection~\cite{Iaia2022}. By directly measuring quasi-instantaneous qubit excitation and de-excitation rates with and without QP injection, we isolate QP-induced energy relaxation from other loss channels. For certain qubit parameters, we observe two peaks in the background-subtracted QP-induced de-excitation rate $\Delta\Gamma_\downarrow$ as a function of the external magnetic flux bias~$\Phi_\mathrm{ext}$. The peaks are centered at the IFQ and HFQ and have similar magnitudes. This observation is inconsistent with simultaneously assuming $\delta\Delta = 0$ and $x_\mathrm{qp}^\mathrm{array} = x_\mathrm{qp}^\mathrm{small}$, which, for our device, would imply \mbox{$\Delta\Gamma_\downarrow(\Phi_\mathrm{ext}=0) \ll \Delta\Gamma_\downarrow(\Phi_\mathrm{ext}=\Phi_0/2)$}, where $\Phi_0$ is the SC flux quantum. We show that accounting for a finite $\delta\Delta$, independently extracted from experiment, reconciles the discrepancy without invoking $x_\mathrm{qp}^\mathrm{array} \neq x_\mathrm{qp}^\mathrm{small}$. Specifically, owing to the condition $\delta E_\mathrm{qp} \lesssim \delta\Delta - \hbar\omega_{01}$, the gap difference suppresses the QP-induced de-excitation rate at the HFQ, making the peak at the HFQ comparable in magnitude to the IFQ peak. The need to account for $\delta\Delta$ therefore provides a plausible explanation for the discrepancy between previously reported bounds on the QP densities in the different circuit elements.

These results underscore the importance of accounting for the hierarchy of the relevant energy scales, $\hbar\omega_{01}$, $\delta\Delta$, and $\delta E_\mathrm{qp}$, when describing QP-induced energy relaxation in low-frequency qubits. They also demonstrate that SC gap asymmetry can strongly suppress QP-induced decoherence in fluxonium, suggesting that deliberate gap engineering may provide an effective mitigation strategy, as it does in transmons.

\begin{figure}
 \includegraphics[width=\columnwidth]{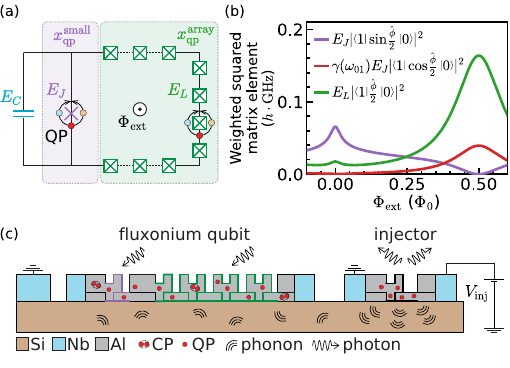}
 \caption{{\bf Experiment concept.} (a)~Circuit diagram of a~fluxonium qubit with the small Josephson junction (purple), the junction array (green), and the capacitor pads (light blue) connected in parallel. An external magnetic flux $\Phi_\mathrm{ext}$ threading the loop formed by the small junction and the junction array controls the SC phase drops across the junctions. QPs are present at two distinct locations: in the small junction (shaded in purple), with the corresponding QP density denoted by $x_\mathrm{qp}^\mathrm{small}$, and in the junction array (shaded in green), with density $x_\mathrm{qp}^\mathrm{array}$. QPs in a~SC are coherent superpositions of electrons (light blue circles) and holes (light orange circles), whose interference upon tunneling determines the QP-induced qubit transition rates. (b)~The functions $E_J\left|\mel{1}{\sin(\hat{\varphi}/2)}{0}\right|^2$, $\gamma(\omega_{01})E_J\left|\mel{1}{\cos(\hat{\varphi}/2)}{0}\right|^2$, and $E_L\left|\mel{1}{\hat{\varphi}/2}{0}\right|^2$ set the relative contributions of the small junction and the junction array to the QP-induced transition rates between the qubit states $\ket{0}$ and $\ket{1}$. The factor $\gamma(\omega_{01})$ accounts for the suppression of the QP tunneling structure factor $S_{-}^\mathrm{small}(\omega_{01})$ relative to $S_{+}^\mathrm{small}(\omega_{01})$. The small-junction contribution dominates at the IFQ ($\Phi_\mathrm{ext} = 0.0 \,\Phi_0$), whereas the junction array contribution dominates at the HFQ ($\Phi_\mathrm{ext} = 0.5\,\Phi_0$). However, the small-junction contribution remains finite at the HFQ because the cosine-term matrix element $\left|\mel{1}{\cos(\hat{\varphi}/2)}{0}\right|^2$ is nonzero. The circuit parameters are $E_C/h = 1.086\,\mathrm{GHz}$, $E_J/h = 2.043\,\mathrm{GHz}$, and $E_L/h = 0.078\,\mathrm{GHz}$. The factor $\gamma(\omega_{01})$ is computed assuming $\delta\Delta/h = 1.72\,\mathrm{GHz}$ and $k_B T_\mathrm{qp}/h = 1.2\,\mathrm{GHz}$. (c)~Cross-section schematic of the experiment showing Josephson junctions in a fluxonium qubit and a QP injector (a voltage-biased Josephson junction) on a~Si substrate. All junctions are Al/AlO$_\mathrm{x}$/Al, while the ground plane and the capacitor pads of the qubits are made of Nb. Relaxation and recombination of QPs in the injector generate CP-breaking phonons ($|V_\mathrm{inj}| > 2\Delta_\mathrm{Al}/e$), which propagate to the qubit region and create QPs in the small junction (purple outline) and in the junction array (green outline). Additionally, Josephson radiation can generate CP-breaking photons ($|V_\mathrm{inj}| > \Delta_\mathrm{Al}/e$), which create QPs in the junction films.}
 \label{fig:overview}
\end{figure}

\section{Quasiparticle-induced qubit transition rates}
\label{sec:qp_induced_rates}

We begin by introducing the circuit Hamiltonian and explaining how the contributions of the small junction and the junction array to the QP-induced transition rates can be disentangled.

The Hamiltonian of the fluxonium circuit~\cite{Manucharyan2009}, shown in Fig.~\ref{fig:overview}(a), is
\begin{equation}
\hat{H} = 4E_C \hat{n}^2 - E_J \cos\hat{\varphi} 
+ \frac{1}{2} E_L\left(\hat{\varphi} - \varphi_\mathrm{ext}\right)^2.
\label{eq:hamiltonian}
\end{equation}
Here, $\hat{\varphi}$ is the phase drop across the small Josephson junction and $\hat{n}$ is the reduced charge operator conjugate to $\hat{\varphi}$, satisfying $[\hat{\varphi}, \hat{n}] = i$. The Hamiltonian includes three energy scales: the Josephson energy $E_J = I_c\Phi_0/(2\pi)$, where $I_c$ is the critical current of the small Josephson junction, the inductive energy $E_L = (\hbar/2e)^2/L$, where $L$ is the superinductance provided by the junction array, and the charging energy $E_C = e^2/(2C)$, where $C$ is the total shunt capacitance. The external flux enters the fluxonium Hamiltonian through the reduced flux $\varphi_\mathrm{ext} = 2\pi \Phi_\mathrm{ext} / \Phi_0$, where $\Phi_\mathrm{ext}$ is the external magnetic flux threading the SC loop, $\Phi_0 = h/(2e)$ is the SC flux quantum, and $-e$ is the electron charge.

QP-induced transitions in fluxonium arise from inelastic QP tunneling across its Josephson junctions, during which energy is exchanged between the qubit and the QPs. These tunneling events can occur in both the small junction and the array junctions. We can separate their respective contributions to the qubit transition rates by using their different dependence on the external flux since, according to the theory, the relative contributions of the QPs in the junction array and in the small junction strongly depend on $\Phi_\mathrm{ext}$.
Specifically, the total transition rate from qubit state~$\ket{i}$ to state~$\ket{f}$ due to QPs is expressed in terms of the flux-dependent matrix elements and transition-frequency-dependent QP structure factors~\cite{Catelani2011, Glazman2021} as
\begin{align}
\label{eq:rates}
\Gamma_{if} = \Bigl|\mel{f}{\frac{\hat{\varphi}}{2}}{i}\Bigr|^2&
\frac{E_L}h S_+^\mathrm{array}(\omega_{if}) \notag\\
+\Bigl|\mel{f}{\sin\frac{\hat{\varphi}}{2}}{i}\Bigr|^2&
\frac{E_J}h S_+^\mathrm{small}(\omega_{if})\notag \\
+\Bigl|\mel{f}{\cos\frac{\hat{\varphi}}{2}}{i}\Bigr|^2&
\frac{E_J}h S_-^\mathrm{small}(\omega_{if})\,.
\end{align}
Here, the term in the first line describes the contribution of the junction array, while the terms in the second and third lines describe the contributions of the small junction, both of which must, in general, be taken into account~\cite{Catelani2012, Azar2026Th}. We define the QP structure factors such that the energy scales corresponding to the small junction $E_J$ and the junction array $E_L$ are factored out. The QP tunneling structure factors $S_{\pm}^\mathrm{small}(\omega_{if})$ and $S_{+}^\mathrm{array}(\omega_{if})$ encode information about the QP density and energy distribution. These dimensionless structure factors depend on the qubit transition energy $\hbar\omega_{if} = E_f - E_i$, the gap difference $\delta \Delta$, and the characteristic width of the QP energy distribution $\delta E_\mathrm{qp}$. The explicit expressions are~\cite{Marchegiani2022, Glazman2021}
\begin{equation}
\label{eq:s_qp}
\begin{split}
&S^{\alpha}_{\pm}(\omega_{if})
=\frac{16}{\bar{\Delta}}\int_{0}^{\infty} 
dE\,f_L^{\alpha}(E)\,D(E,\Delta_L) \\
&\times
D(E-\hbar\omega_{if},\Delta_H)\Bigl(1  \pm \tfrac{\Delta_L \Delta_H}{E(E-\hbar\omega_{if})}\Bigr)
+ (L \leftrightarrow H).
\end{split}
\end{equation}
Here, $\alpha \in \{\mathrm{small},\mathrm{array}\}$ labels the qubit circuit element, $\bar{\Delta} = (\Delta_L + \Delta_H) / 2$, $f_L^\alpha(E)$ is the QP distribution function in the low-gap lead of the circuit element $\alpha$, and $D(E,\Delta) = E\,\Theta(E-\Delta)/\sqrt{E^2-\Delta^2}$ is the normalized Bardeen--Cooper--Schrieffer (BCS) density of states, where $\Theta(E)$ is the Heaviside step function. The term $(L \leftrightarrow H)$ represents the complementary tunneling contribution from the $\Delta_H$ lead to the $\Delta_L$ lead. We note that $S^\alpha_{-} \ll S_{+}^\alpha$ when the characteristic energies $\hbar\omega_{if}$, $\delta \Delta$, and $\delta E_\mathrm{qp}$ are small compared to $\Delta_L$ and $\Delta_H$~[Appendix~\ref{app:qp_rate_gen_model}].

In this work, we focus on transitions between the two lowest-energy qubit states $\ket{0}$ and $\ket{1}$. Figure~\ref{fig:overview}(b) shows the external flux dependence of appropriately scaled matrix elements, illustrating the relative contributions of QP tunneling events in the small junction and the junction array to the qubit transition rates. This comparison assumes equal QP tunneling structure factors $S_{+}^\mathrm{small}(\omega_{01}) = S_{+}^\mathrm{array}(\omega_{01})$ and accounts for the suppression of the cosine-term contribution through $S_{-}^\mathrm{small}(\omega_{01}) = \gamma(\omega_{01})\,S_{+}^\mathrm{small}(\omega_{01})$, where $\gamma(\omega_{01}) \ll 1$~\cite{Glazman2021, Azar2026Th}.
The small-junction contribution dominates at the IFQ, whereas the junction array contribution dominates at the HFQ. Indeed, the sine-term contribution of the small junction vanishes at the HFQ, since $\left|\mel{1}{\sin (\hat{\varphi}/2)}{0}\right|^2 = 0$~\cite{Pop2014, Glazman2021}, leaving only a subleading cosine-term contribution [Fig.~\ref{fig:overview}(b)]. By contrast, the array contribution is maximal at the HFQ and relatively small at the IFQ. By measuring the QP-induced increase in the excitation $\Gamma_\uparrow \equiv \Gamma_{01}$ and de-excitation $\Gamma_\downarrow \equiv \Gamma_{10}$ transition rates and taking advantage of their expected dependence on $\Phi_\mathrm{ext}$, the contributions from the small junction and the junction array can be disentangled with minimal assumptions about the QP energy distribution function $f(E)$, which determines $S_{\pm}^\mathrm{small}(\omega)$ and $S_{+}^\mathrm{array}(\omega)$. This, in turn, allows for a direct evaluation of the corresponding QP densities, $x_\mathrm{qp}^\mathrm{small}$ and $x_\mathrm{qp}^\mathrm{array}$.

\section{Experiment}
\label{sec:experiment}

\begin{figure}
 \includegraphics[width=\columnwidth]{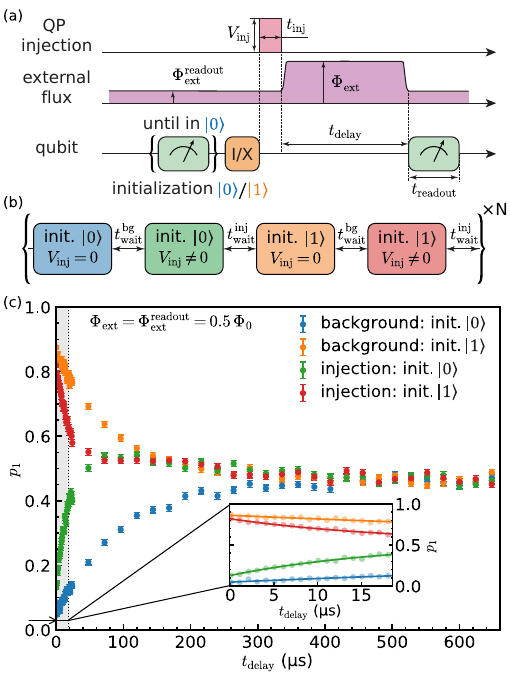}
 \caption{{\bf Transition rate measurements.} (a) Elementary pulse sequence for measuring the qubit transition rates. The qubit is initialized in either the ground state $\ket{0}$ or the excited state $\ket{1}$ at the readout flux $\Phi_\mathrm{ext}^\mathrm{readout}$ by preselecting the qubit in the $\ket{0}$ state and applying either an $I$ or $X$ gate. $\Phi_\mathrm{ext}^\mathrm{readout} = 0.5\,\Phi_0$ for the data presented in the main text. A QP injection pulse is applied after initialization, when appropriate, immediately followed by tuning the external flux to the target value $\Phi_\mathrm{ext}^\mathrm{target} \equiv \Phi_\mathrm{ext}$, where the qubit idles for a time $t_\mathrm{delay}$. The external flux is then returned to $\Phi_\mathrm{ext}^\mathrm{readout}$ for readout of the final qubit state. (b)~Sequence of interleaved elementary measurements, in which each block represents an elementary measurement shown in panel (a). With the qubit initialized in the $\ket{0}$ state, a~pair of measurements separated by $t_\mathrm{wait}^\mathrm{bg} \ge 0.8\,\mathrm{ms}$ is performed without an~injection pulse (blue block) and with an~injection pulse (green block). After a longer wait time, $t_\mathrm{wait}^\mathrm{inj} \ge 16\,\mathrm{ms}$, the same pair of measurements is performed for the initial state $\ket{1}$ (orange and red blocks). The sequence is repeated $N$ times. (c)~Excited state population $p_1$ as a function of $t_\mathrm{delay}$. Blue and orange symbols (for the qubit prepared in the $\ket{0}$ and $\ket{1}$ states, respectively) correspond to background measurements with $V_\mathrm{inj} = 0$, while green and red symbols (also for the qubit prepared in the $\ket{0}$ and $\ket{1}$ states) correspond to the QP injection measurements with $V_\mathrm{inj} = 7.0 \,\Delta_\mathrm{Al}/e$ and $t_\mathrm{inj} = 2.0 \,\mu\mathrm{s}$. For this dataset, $N = 2400$ and $t_\mathrm{readout} = 5.4 \,\mu\mathrm{s}$. The inset zooms in on an 18-$\mu\mathrm{s}$ interval, shaded in gray in the main plot. The de-excitation $\Gamma_\downarrow$ and excitation $\Gamma_\uparrow$ transition rates are obtained from local joint fits (solid lines) applied separately to the background (blue and orange) and injection (green and red) data. The extracted transition rates are $\Gamma_\downarrow^\mathrm{bg} = 6.3 \,\mathrm{ms^{-1}}$, $\Gamma_\uparrow^\mathrm{bg} = 5.2 \,\mathrm{ms^{-1}}$, $\Gamma_\downarrow^\mathrm{inj} = 25.4 \,\mathrm{ms^{-1}}$, and $\Gamma_\uparrow^\mathrm{inj} = 28.8 \,\mathrm{ms^{-1}}$.}
 \label{fig:protocol}
\end{figure}

\subsection{Experiment design}
\label{sec:exp_design}
Each $10\,\mathrm{mm} \!\times\! 10\,\mathrm{mm}$ chip measured in this work includes fluxonium and transmon qubits, together with injector structures, co-fabricated on a~pre-cleaned high-resistivity Si substrate. The small Josephson junctions of the qubits, the junctions in the inductor arrays, and the injectors are Al/AlO$_\mathrm{x}$/Al junctions fabricated using the Dolan-bridge technique~\cite{Dolan1977}. The qubit capacitor pads, coplanar waveguide resonators, and the transmission line for readout, as well as the qubit control and flux bias lines, are all patterned in the Nb ground plane layer. The fabrication is detailed in Appendix~\ref{app:fab}. Measurements are performed in a dilution refrigerator with a base temperature below $10~\mathrm{mK}$ [Appendix~\ref{app:exp_setup}], and the extracted qubit device parameters are summarized in Appendix~\ref{app:qubits}. For clarity, in the main text we present data only for one fluxonium device, referred to as qubit Q1 in the Appendices.

To enable controlled QP poisoning, a dedicated Al/AlO$_\mathrm{x}$/Al Josephson junction injector embedded between two Nb pads is co-fabricated with the qubits on the same chip [Fig.~\ref{fig:overview}(c)]. Applying a voltage bias~$V_\mathrm{inj}$ above $2\Delta_\mathrm{Al}/e$, where $\Delta_\mathrm{Al}$ is the SC gap of Al, drives a QP current. The subsequent relaxation of QPs toward the gap edge and their eventual recombination near the injector generate CP-breaking phonons that propagate through the Si substrate to the qubit region~\cite{Iaia2022, Yelton2024}. Upon reaching the qubit region, these athermal phonons break CPs in the Al junction leads, thereby increasing the QP density in the qubit junctions. For the short injection pulses used in the main experiments, QP poisoning via CP-breaking phonons dominates over the poisoning via CP-breaking photons from Josephson radiation emitted when $|V_\mathrm{inj}|>\Delta_\mathrm{Al}/e$ [Appendix~\ref{app:photon_poisoning}].
 
\subsection{Measurement of the qubit transition rates}
\label{sec:meas_protocol}

To characterize QP-induced qubit relaxation, we go beyond conventional $T_1$ measurements. The QP-induced rates after injection are time-dependent as the QP density begins to recover toward its background value~[Appendix~\ref{app:t1_recovery}]. Measurement of transition rates using the conventional $T_1$ protocol, which relies on fitting an exponential decay over a time window comparable to the QP-density recovery time, convolutes the qubit relaxation dynamics with the QP recovery process. Moreover, measuring population decay under QP injection can lead to a subtle illusory increase in $T_1$~\cite{Gustavsson2016, Spiecker2023}.

To overcome these limitations, we analyze the qubit population dynamics on short timescales. When the population dynamics is restricted to the two lowest qubit levels, $p_0(t) + p_1(t) = 1$ and $\dot{p}_1(t) = -\Gamma_\downarrow(t) p_1(t) + \Gamma_\uparrow(t) p_0(t)$, where $p_0(t)$ and $p_1(t)$ are qubit populations in the ground and excited states at time $t$. Under the assumption of approximately constant rates within a short time window, we can measure the qubit population $p_1(t)$ starting with two different initial conditions, preferably $p_1(0)=0$ and $p_1(0)=1$, which maximize the separation between the two population traces and provide sensitivity to both the excitation and de-excitation rates. We then perform a joint fit to the two curves to extract the quasi-instantaneous rates $\Gamma_\downarrow$ and $\Gamma_\uparrow$ and the initial qubit populations $p_1^{\ket{0}}(0)$ and $p_1^{\ket{1}}(0)$, where the superscripts label the two preferred initial conditions. The joint fit of two curves yields four parameters, allowing for a robust determination of the $\Gamma_\downarrow$ and $\Gamma_\uparrow$ rates while accounting for state-preparation infidelities. We note that this model neglects the possibility of qubit leakage into non-computational states; we discuss the role of leakage in detail in Appendices~\ref{app:three_level_rate_model} and~\ref{app:rate_inversion}.

Figure~\ref{fig:protocol} shows our measurement protocol and the representative population curves obtained using it. Each elementary experiment presented in Fig.~\ref{fig:protocol}(a) begins by preparing the qubit in either the ground state $\ket{0}$ or the excited state $\ket{1}$ at a~chosen readout flux $\Phi_\mathrm{ext}^\mathrm{readout}$. We take advantage of the high tunability of fluxonium to select the $\Phi_\mathrm{ext}^\mathrm{readout}$ point with a large dispersive shift $\chi_{01} \equiv \chi_1-\chi_0$ for better state discrimination. For all data presented in the main text, $\Phi_\mathrm{ext}^\mathrm{readout} = 0.5\,\Phi_0$. The qubit is initialized by repeatedly measuring it until it is found in the $\ket{0}$ state with a $100$-$\mu\mathrm{s}$ delay between repetitions, followed by an $I$ or $X$ gate to prepare the desired initial state. After initialization, an injection pulse $V_\mathrm{inj}$ is applied for a duration $t_\mathrm{inj}$; for background measurements, $V_\mathrm{inj}$ is set to zero. The external flux is then shifted to the target value $\Phi_\mathrm{ext}$ adiabatically with respect to the qubit-level dynamics, yet rapidly compared with the qubit-relaxation and QP-recovery timescales. The qubit idles at $\Phi_\mathrm{ext}$ for a variable delay time $t_\mathrm{delay}$, after which the flux is tuned back to $\Phi_\mathrm{ext}^\mathrm{readout}$ for readout. Such dynamic flux biasing is commonly employed in swap spectroscopy measurements, where the qubit is tuned between different flux bias points~\cite{Lisenfeld2016, Bilmes2022, Sun2023}. The elementary measurements, with and without QP injection, are interleaved as shown in Fig.~\ref{fig:protocol}(b) to minimize the effects of drift on the collected statistics.

\begin{figure}
 \includegraphics[width=\columnwidth]{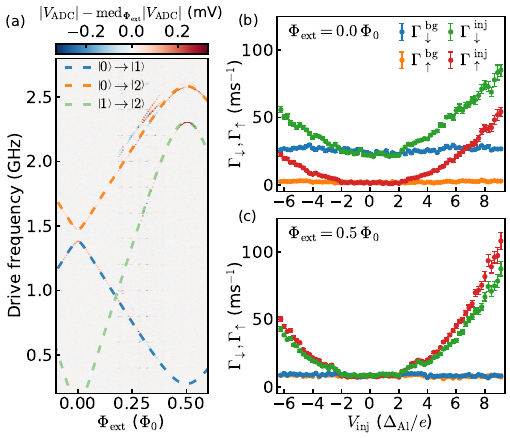}
 \caption{{\bf Qubit spectroscopy and dependence of the transition rates on the injection bias.}
 (a)~Conventional two-tone qubit spectroscopy as a function of $\Phi_\mathrm{ext}$. The blue, orange, and green dashed lines are the $\ket{0}\!\to\!\ket{1}$, $\ket{0}\!\to\!\ket{2}$, and $\ket{1}\!\to\!\ket{2}$ qubit transitions obtained from a fit to the measured spectrum. The extracted qubit parameters are $E_C/h = 1.086\,\mathrm{GHz}$, $E_J/h = 2.043\,\mathrm{GHz}$, and $E_L/h = 0.078\,\mathrm{GHz}$. (b,\,c)~Measured background transition rates, $\Gamma^\mathrm{bg}_\downarrow$ (blue) and $\Gamma^\mathrm{bg}_\uparrow$ (orange), and post-injection transition rates, $\Gamma^\mathrm{inj}_\downarrow$ (green) and $\Gamma^\mathrm{inj}_\uparrow$ (red), at the IFQ (b) and HFQ (c), plotted as a function of the injection bias $V_\mathrm{inj}$. For both datasets, $t_\mathrm{inj} = 4.0 \,\mu\mathrm{s}$. The transition rates are extracted from joint fits to the population data in the $0 \leq t_\mathrm{delay} \leq 16 \,\mu\mathrm{s}$ time window.}
 \label{fig:rates_inj_bias}
\end{figure}

\subsection{Injection bias dependence of the qubit transition rates}
\label{sec:rates_vs_inj_bias}

Using the measurement protocol described above, we characterize the response of the qubit to QP injection by measuring the qubit transition rates as a function of injection bias $V_\mathrm{inj}$ at the IFQ and HFQ flux bias points, identified using conventional two-tone qubit spectroscopy [Fig.~\ref{fig:rates_inj_bias}(a)]. The measured background (no injection) $\Gamma^\mathrm{bg}_{\downarrow\,,\uparrow}$ and post-injection $\Gamma^\mathrm{inj}_{\downarrow\,,\uparrow}$ transition rates are shown in Fig.~\ref{fig:rates_inj_bias}(b) for the IFQ and in Fig.~\ref{fig:rates_inj_bias}(c) for the HFQ.

At both the IFQ and HFQ, the background rates $\Gamma^\mathrm{bg}_\downarrow$ and $\Gamma^\mathrm{bg}_\uparrow$ remain mostly constant, with minor variations attributed to slow drifts of the qubit relaxation time. We note that $\Gamma^\mathrm{bg}_\uparrow \ll \Gamma^\mathrm{bg}_\downarrow$ at $\Phi_\mathrm{ext} = 0.0 \,\Phi_0$, whereas $\Gamma^\mathrm{bg}_\uparrow \lesssim \Gamma^\mathrm{bg}_\downarrow$ at $\Phi_\mathrm{ext} = 0.5\,\Phi_0$. Using the detailed-balance relation $\Gamma^\mathrm{bg}_\uparrow / \Gamma^\mathrm{bg}_\downarrow = \exp(-\hbar\omega_{01} / k_B T_q)$, where $k_B$ is the Boltzmann constant, we estimate the effective qubit temperature $T_q$ to range from approximately $35 \,\mathrm{mK}$ at $\Phi_\mathrm{ext} = 0.0 \,\Phi_0$ to $80 \,\mathrm{mK}$ at $\Phi_\mathrm{ext} = 0.5\,\Phi_0$. We speculate that the elevation of the effective qubit temperature above the mixing-chamber temperature and its dependence on external flux may result from insufficient filtering of infrared radiation~\cite{Houzet2019, Diamond2022, Connolly2024}.

In the injection datasets of Figs.~\ref{fig:rates_inj_bias}(b,\,c), the post-injection rates match the background values for $\left|V_\mathrm{inj}\right| < 2\Delta_\mathrm{Al}/e$, as expected. In contrast, for $\left|V_\mathrm{inj}\right| > 2\Delta_\mathrm{Al}/e$, the transition rates increase above the background level, $\Gamma^\mathrm{inj}_{\downarrow(\uparrow)} > \Gamma^\mathrm{bg}_{\downarrow(\uparrow)}$, indicating an enhancement of the transition rates due to QP poisoning. The well-defined onset at $2\Delta_\mathrm{Al}/e$ is associated with the generation of CP-breaking phonons at the injector junction, which in turn propagate through the substrate and break CPs in the qubit junctions, thereby increasing the QP density. We note that the calibration between the voltage applied at the input of the injection line and the chip-level voltage $V_\mathrm{inj}$ was performed independently before the cooldown. The agreement between the onset of the post-injection rate enhancement and the calibrated value of $2\Delta_\mathrm{Al}/e$ is consistent with phonon-mediated poisoning above the CP-breaking threshold. Notably, at the IFQ the post-injection transition rates exhibit regular ordering, $\Gamma^\mathrm{inj}_\uparrow < \Gamma^\mathrm{inj}_\downarrow$, whereas at the HFQ they demonstrate an~apparent inverse ordering, $\Gamma^\mathrm{inj}_\uparrow > \Gamma^\mathrm{inj}_\downarrow$, in the measured range of $V_\mathrm{inj}$ above $2\Delta_\mathrm{Al}/e$. We discuss the origin of this apparent QP-induced rate inversion in Sec.~\ref{sec:excitation_rate}.

\subsection{External flux dependence of the qubit transition rates}
\label{sec:rates_vs_ext_flux}

\begin{figure}
 \includegraphics[width=\columnwidth]{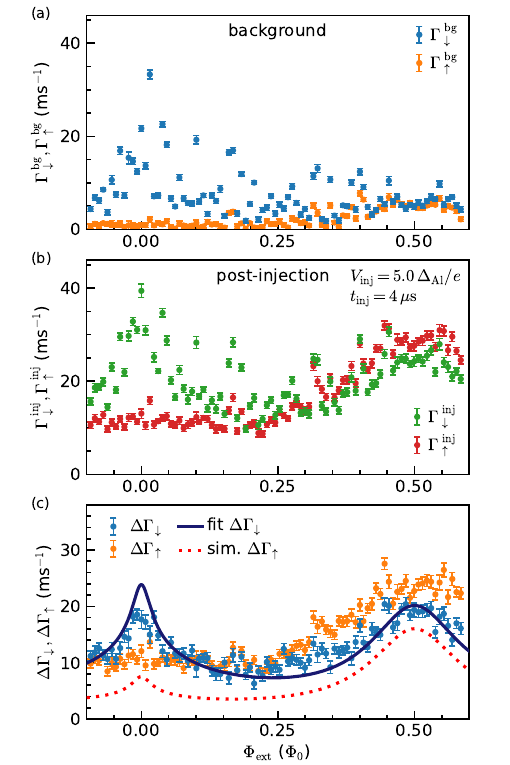}
 \caption{{\bf Dependence of the transition rates on external flux.} (a)~Background de-excitation $\Gamma_\downarrow^\mathrm{bg}$ (blue) and excitation $\Gamma_\uparrow^\mathrm{bg}$ (orange) rates as functions of external flux~$\Phi_\mathrm{ext}$. (b)~Post-injection de-excitation $\Gamma_\downarrow^\mathrm{inj}$ (green) and excitation $\Gamma_\uparrow^\mathrm{inj}$ (red) rates following an injection pulse with $V_\mathrm{inj} = 5.0 \,\Delta_\mathrm{Al}/e$. Other measurement parameters are as in Fig.~\ref{fig:rates_inj_bias}. (c)~QP-induced increase in the de-excitation ($\Delta\Gamma_\downarrow = \Gamma_\downarrow^\mathrm{inj} - \Gamma_\downarrow^\mathrm{bg}$) and excitation ($\Delta\Gamma_\uparrow = \Gamma_\uparrow^\mathrm{inj} - \Gamma_\uparrow^\mathrm{bg}$) transition rates as functions of external flux~$\Phi_\mathrm{ext}$. The measured rates exhibit an apparent inversion, $\Delta\Gamma_\uparrow > \Delta\Gamma_\downarrow$, near the HFQ, while retaining their regular ordering, $\Delta\Gamma_\uparrow < \Delta\Gamma_\downarrow$, at the IFQ. The blue solid curve is a one-parameter fit to $\Delta\Gamma_\downarrow$ with the common QP density $x_\mathrm{qp}$ as the free parameter. The red dotted curve is the corresponding prediction for~$\Delta\Gamma_\uparrow$. The independently determined parameters are $\delta\Delta/h = 1.72\,\mathrm{GHz}$ and $k_B T_\mathrm{qp}/h = 1.2\,\mathrm{GHz}$. The discrepancy in $\Delta\Gamma_\uparrow$  between the theory and the data is likely due to leakage to higher qubit states, as discussed in Sec.~\ref{sec:excitation_rate} and Appendix~\ref{app:leakage_to_noncomp_states}.}
 \label{fig:rates_ext_flux}
\end{figure}

Figure~\ref{fig:rates_ext_flux}(a) shows the background transition rates $\Gamma_\downarrow^\mathrm{bg}(\Phi_\mathrm{ext})$ and $\Gamma_\uparrow^\mathrm{bg}(\Phi_\mathrm{ext})$, while Figure~\ref{fig:rates_ext_flux}(b) shows the post-injection transition rates $\Gamma_\downarrow^\mathrm{inj}(\Phi_\mathrm{ext})$ and $\Gamma_\uparrow^\mathrm{inj}(\Phi_\mathrm{ext})$ at $V_\mathrm{inj} = 5.0\,\Delta_\mathrm{Al}/e$. Over the entire range of the external flux, we observe $\Gamma^\mathrm{inj}_{\downarrow(\uparrow)}(\Phi_\mathrm{ext}) > \Gamma^\mathrm{bg}_{\downarrow(\uparrow)}(\Phi_\mathrm{ext})$, consistent with the injection-induced increase in the transition rates. As in Fig.~\ref{fig:rates_inj_bias}(c), the excitation and de-excitation rates exhibit an apparent inversion near the HFQ after injection. Both background and post-injection transition rates exhibit pronounced jumps attributed to two-level systems (TLSs) that are coupled to the qubit at various flux bias points~\cite{Sun2023, Zhuang2026}.

To isolate the QP-induced contribution to the transition rates, we subtract the background rates from the post-injection rates to obtain the background-subtracted transition rates $\Delta\Gamma_\downarrow = \Gamma_\downarrow^\mathrm{inj} - \Gamma_\downarrow^\mathrm{bg}$ and $\Delta\Gamma_\uparrow = \Gamma_\uparrow^\mathrm{inj} - \Gamma_\uparrow^\mathrm{bg}$, which are shown in Fig.~\ref{fig:rates_ext_flux}(c). In contrast to the individual background and post-injection rates, these background-subtracted (QP-induced) rates do not exhibit TLS jumps and show a~smoother dependence on the external flux.

The flux dependence of the QP-induced de-excitation rate $\Delta\Gamma_\downarrow$ (blue data points in Fig.~\ref{fig:rates_ext_flux}(c)) exhibits two peaks of comparable magnitude, centered at the IFQ and HFQ. The QP-induced excitation rate $\Delta\Gamma_\uparrow$ (orange data points) has a smoother flux dependence; in contrast to the de-excitation rate, the excitation rate at the HFQ appreciably exceeds that at the IFQ. Notably, in a range of fluxes around the HFQ, the rate of QP-induced excitation exceeds the rate of QP-induced de-excitation. This behavior contradicts the na\"ive detailed balance assumption, even for an elevated temperature. We discuss its possible origins in Sec.~\ref{sec:excitation_rate}.

The nontrivial external flux dependence of the QP-induced qubit transition rates $\Delta\Gamma_\uparrow$ and $\Delta\Gamma_\downarrow$ calls for the detailed analysis presented in the next section.

\section{Comparison with theory}
\label{sec:comp_with_theory}

Two assumptions commonly made~\cite{Catelani2011, Glazman2021} when describing QP-induced dissipation are that (i)~the SC gaps in the junction leads are equal, $\Delta_L = \Delta_H$, and that (ii)~the~characteristic width of the QP energy distribution is much smaller than the qubit transition energy between the ground and first excited states, $\delta E_\mathrm{qp} \ll \hbar\omega_{01}$. In principle, as shown in Appendix~\ref{app:simple_model} and Table~\ref{tab:xqp_models}, the \textit{de-excitation} data can be explained under these assumptions by taking $x_\mathrm{qp}^\mathrm{array} \ll x_\mathrm{qp}^\mathrm{small}$. We begin by critically examining the validity of assumptions (i) and (ii) for our injection experiment, starting with the latter.

\subsection{Quasiparticle energy distribution}
\label{sec:qp_energy_distribution}
As we now show, assumption $\delta E_\mathrm{qp} \ll \hbar\omega_{01}$ must be ruled out. Indeed, if this assumption were valid, QPs would not have sufficient energy to excite the qubit, leading to $\Delta\Gamma_\uparrow \simeq 0$. This prediction is in stark disagreement with the experimental data, which show a finite $\Delta\Gamma_\uparrow$ comparable with $\Delta\Gamma_\downarrow$ throughout the external flux range. This indicates that $\delta E_\mathrm{qp} \sim \hbar\omega_{01}$ in our experiment.

From the injection response of a transmon qubit co-fabricated on the same chip, with a transition frequency $\omega_{01}^\mathrm{tr}/2\pi \approx 3.81\,\mathrm{GHz}$ [Appendix~\ref{app:qubits_and_transmons}], we measure the QP-induced transition rates $\Delta\Gamma_\uparrow^\mathrm{tr} = 3\,\mathrm{ms}^{-1}$ and $\Delta\Gamma_\downarrow^\mathrm{tr} = 81\,\mathrm{ms}^{-1}$ under the same injection conditions as those used for the fluxonium dataset in Fig.~\ref{fig:rates_ext_flux}. Using the detailed-balance relation $\Delta\Gamma_\uparrow^\mathrm{tr} / \Delta\Gamma_\downarrow^\mathrm{tr} = e^{-\hbar\omega_{01}^\mathrm{tr} / k_B T_\mathrm{qp}}$, we estimate the effective QP temperature to be $k_B T_\mathrm{qp}/h \approx 1.2\,\mathrm{GHz}$. This implies that the approximation $\delta E_\mathrm{qp} \ll \hbar\omega_{01}$ is indeed violated for the fluxonium qubit, for which $\omega_{01}/2\pi \approx 0.28\,\mathrm{GHz}$ at the HFQ and $\omega_{01}/2\pi \approx 1.38\,\mathrm{GHz}$ at the IFQ.

We can relax the assumption $\delta E_\mathrm{qp}\ll\hbar\omega_{01}$ by introducing a thermal QP energy distribution with an effective temperature $T_\mathrm{qp}$:
\begin{equation}
f_{L,H}^{\alpha}(E) = 
x_\mathrm{qp}^\alpha \sqrt{\frac{\Delta_L}{2\pi k_B T_\mathrm{qp}}}
e^{-\frac{E-\Delta_L}{k_B T_\mathrm{qp}}}.
\label{eq:f_th}
\end{equation}
Here, $x_\mathrm{qp}^\alpha$ with $\alpha \in \{\mathrm{small},\mathrm{array}\}$ denotes the QP density on the low-gap side of the corresponding circuit element. Introducing $x_\mathrm{qp}^\alpha$ allows the QP density to remain out of equilibrium while the energy distribution is still characterized by an effective temperature $T_\mathrm{qp}$~\cite{Connolly2024, Marchegiani2025}.

\subsection{Superconducting gap asymmetry}
\label{sec:sc_gap_asymmetry}

If $\Delta_L = \Delta_H$ and the QP densities in the junction array and the small junction are set equal, $x_\mathrm{qp}^\mathrm{array} = x_\mathrm{qp}^\mathrm{small}$, Eqs.~\eqref{eq:rates},~\eqref{eq:s_qp}, and~\eqref{eq:f_th} predict $\Delta\Gamma_\downarrow(\Phi_0/2) \,/ \,\Delta\Gamma_\downarrow(0) \approx 3.9$, implying that the HFQ peak should substantially exceed the IFQ peak. This prediction is in clear disagreement with the data in Fig.~\ref{fig:rates_ext_flux}(c), where the two peaks have similar heights. This discrepancy points to the failure of the model with $\Delta_L = \Delta_H$ and $x_\mathrm{qp}^\mathrm{array} = x_\mathrm{qp}^\mathrm{small}$, although it does not identify which assumption fails. As we now explain, a finite SC gap difference $\Delta_L \neq \Delta_H$ is expected in our sample, and there is insufficient evidence in our data to relax the assumption $x_\mathrm{qp}^\mathrm{array} = x_\mathrm{qp}^\mathrm{small}$.

The SC gap of aluminum depends strongly on film thickness. In our devices, the thicknesses of the two Al films that form the Josephson junction leads are 40 and 80\,nm, corresponding to an estimated gap difference $\delta\Delta/h \approx 1.8\,\mathrm{GHz}$ [Appendix~\ref{app:qubits}]. To refine this estimate, we measure the QP-induced $\ket{0}\!\to\!\ket{2}$ transition rate, which is expected to be resonantly enhanced at the flux bias where the qubit transition energy matches the SC gap difference, $\hbar\omega_{02} = \delta\Delta$~\cite{Diamond2022, Marchegiani2022}. We observe a peak at $\omega_{02}/2\pi \approx 1.72\,\mathrm{GHz}$ [Appendix~\ref{app:three_level_rates_vs_flux}], consistent with the thickness-based estimate. We therefore use $\delta\Delta/h = 1.72\,\mathrm{GHz}$ in the analysis that follows.

\subsection{Quasiparticle-induced de-excitation rate}
\label{sec:de-excitation_asymmetric_case}

Having independently determined $\delta\Delta$ and $T_\mathrm{qp}$, we now fit the $\Delta\Gamma_\downarrow$ data in Fig.~\ref{fig:rates_ext_flux}(c). We impose $x_\mathrm{qp}^\mathrm{array} = x_\mathrm{qp}^\mathrm{small}$ and thus use the common QP density $x_\mathrm{qp}$ as the only free fit parameter. This one-parameter fit reproduces the measured flux dependence of $\Delta\Gamma_\downarrow$, including the comparable heights of the two peaks at the IFQ and the HFQ, as shown by the solid blue curve in Fig.~\ref{fig:rates_ext_flux}(c). This agreement reflects the suppression of QP-induced de-excitation at the HFQ by the SC gap difference, owing to the condition \mbox{$k_B T_\mathrm{qp} \lesssim \delta\Delta - \hbar\omega_{01}$}~\cite{Connolly2024, KurilovichV2026}. The extracted QP density $x_\mathrm{qp} = 3.6\times10^{-6}$ is in good agreement with the value independently obtained from the transmon data $x_\mathrm{qp}^\mathrm{tr} = 3.0 \times 10^{-6}$ [Appendix~\ref{app:asymmetric_gap_model}].

To investigate the possibility of unequal QP densities in the small junction and in the junction array, we relax the constraint of equal QP densities and treat the ratio $x_\mathrm{qp}^\mathrm{array}/x_\mathrm{qp}^\mathrm{small}$ as a second fit parameter. The fit (not shown) is visually similar to the one with only one fit parameter. The resulting value $x_\mathrm{qp}^\mathrm{array} / x_\mathrm{qp}^\mathrm{small} = 1.2$ indicates that there is no appreciable imbalance between the QP densities in the two circuit elements.

\subsection{Quasiparticle-induced excitation rate}
\label{sec:excitation_rate}

We now turn to the QP-induced excitation rate $\Delta\Gamma_\uparrow$. The red dotted curve in Fig.~\ref{fig:rates_ext_flux}(c) shows the prediction of Eqs.~\eqref{eq:rates} and~\eqref{eq:s_qp} with the QP distribution given by Eq.~\eqref{eq:f_th}. The calculation uses the experimentally determined $\delta\Delta$ and $T_\mathrm{qp}$, and the QP density extracted from the fit to $\Delta\Gamma_\downarrow$, assuming $\Delta_L \neq \Delta_H$ and $x_\mathrm{qp}^\mathrm{array} = x_\mathrm{qp}^\mathrm{small}$, as discussed above. Thus, the theoretical $\Delta\Gamma_\uparrow$ curve contains no adjustable parameters. Nevertheless, it qualitatively reproduces the measured external flux dependence, including the two-peak structure near the IFQ and the HFQ and the appreciably larger excitation rate at the HFQ than at the IFQ. However, a quantitative disagreement is apparent: the measured $\Delta\Gamma_\uparrow$ systematically exceeds the model prediction. Moreover, in a range of external flux around the HFQ, the measured excitation rate exceeds the de-excitation rate. The latter observation cannot be captured by any thermal QP distribution, for which detailed balance requires $\Delta\Gamma_\uparrow / \Delta\Gamma_\downarrow = e^{-\hbar\omega_{01}/k_B T_\mathrm{qp}} < 1$.

We believe that the apparent excitation rate $\Delta\Gamma_\uparrow$ includes a contribution from leakage to higher qubit states that are misidentified as $\ket{1}$ during readout. Indeed, the estimated QP energy scale, $k_B T_\mathrm{qp}/h \approx 1.2\,\mathrm{GHz}$, is comparable to the fluxonium energy level spacings [Fig.~\ref{fig:spectra}(a)]. Therefore, the theory predicts QP-induced transition rates from $\ket{0}$ to $\ket{2}$, $\ket{3}$, and $\ket{4}$ that are comparable to the $\ket{0}\!\to\!\ket{1}$ rate [Appendix~\ref{app:leakage_to_noncomp_states}]. However, among the leakage states, our readout resolves only the $\ket{2}$ state: the dispersive shifts of $\ket{3}$, $\ket{4}$, $\ket{5}$, and $\ket{6}$ are close to that of $\ket{1}$ [Appendix~\ref{app:leakage_to_noncomp_states}, Fig.~\ref{fig:q1_chi_shifts}], so these states are misidentified as $\ket{1}$. Therefore, the apparent $\Delta\Gamma_\uparrow$ transition rate obtained with the procedure of Sec.~\ref{sec:meas_protocol} may conflate several transitions. A~quantitative description of this apparent rate would require additional parameters to account for state-assignment infidelities and deviations of the QP distribution from a Boltzmann form~\cite{Martinis2009, Marchegiani2025, KurilovichV2026}, which is beyond the scope of the present work. 

By contrast, the extracted de-excitation rate $\Delta\Gamma_\downarrow$ is largely insensitive to these artifacts because the readout clearly separates $\ket{0}$ from all excited states and therefore reliably captures population transfer to $\ket{0}$ when the qubit is initialized in $\ket{1}$. Importantly, the $\ket{1}$ state is prepared by preselecting the qubit in $\ket{0}$ and applying an $X$ gate, rather than by directly preselecting it in $\ket{1}$. Furthermore, because the dispersive shifts of $\ket{3}$ and higher states are close to that of $\ket{1}$, transitions from $\ket{1}$ into these states leave the apparent population $p_1(t)$ unchanged and thus do not appreciably affect the extracted de-excitation rate $\Delta\Gamma_\downarrow$. Finally, we note that revisiting the rate-extraction analysis to explicitly account for transitions involving $\ket{2}$ has only a minor effect on the extracted apparent rates $\Delta\Gamma_\downarrow$ and $\Delta\Gamma_\uparrow$ [Appendix~\ref{app:three_level_rates_vs_flux}, Fig.~\ref{fig:q1_three_level_rates_vs_flux}(a)].

As an alternative interpretation, the apparent rate inversion could arise from a genuinely non-equilibrium QP distribution established following injection. In Appendix~\ref{app:alternative_qp_dist}, we show that reasonable agreement with the data can be obtained by assuming that the effective QP chemical potential on the high-gap side of Josephson junctions is elevated relative to that on the low-gap side~\cite{Marchegiani2025}. However, this model requires a rather narrow QP energy distribution above the gap edge, $\delta E_\mathrm{qp}/h \simeq 0.2\,\mathrm{GHz}$, corresponding to an energy scale comparable to the temperature of the mixing chamber. Relaxation of the injected QPs to such low energies is expected to occur on timescales exceeding milliseconds~\cite{Glazman2021}, far longer than the timescale probed in our measurements. Moreover, this interpretation is inconsistent with the broader width of the QP energy distribution $\delta E_\mathrm{qp}/h \simeq 1.2\,\mathrm{GHz}$ inferred from the injection response of the transmon qubit. We therefore consider leakage into higher qubit states to be the more plausible scenario affecting $\Delta\Gamma_\uparrow$.

\section{Summary and conclusions}
\label{sec:summary}

We investigated QP-induced energy relaxation in fluxonium qubits under controlled on-chip QP injection. Although the insensitivity of fluxonium to offset charge precludes charge-parity detection of QP tunneling, the combination of controlled injection with measurements of quasi-instantaneous transition rates allowed us to isolate the QP-induced contribution to the qubit dynamics.

Our central finding is that the SC gap difference between junction leads $\delta\Delta$, which routinely arises in junction fabrication processes, plays an essential role in QP-induced transitions in fluxonium. The gap difference strongly suppresses the transition rates at flux biases where the qubit transition energy is small compared to $\delta\Delta$, in particular near the HFQ, where fluxonium is commonly operated. This effect may therefore be relevant for the interpretation of the QP-induced dissipative response of fluxonium qubits; see, e.g., Ref.~\cite{Pop2014}.

After accounting for $\delta\Delta$, we are able to describe the full external flux dependence of the qubit de-excitation rates using a model with equal QP densities in the small junction and the junction array, $x_\mathrm{qp}^\mathrm{array} = x_\mathrm{qp}^\mathrm{small}$. This result deserves a comment given the previously reported bounds on QP densities, which are consistently found to be $x_\mathrm{qp}^\mathrm{array} \ll x_\mathrm{qp}^\mathrm{small}$ [Table~\ref{tab:qp_density_summary} in Appendix~\ref{app:xqp_literature}]. We conjecture that this disparity arises in part from analyzing the data with a model that assumes $\delta\Delta = 0$: such a model predicts a particularly strong sensitivity to QPs in the array (due to the strength of the respective transition matrix element near the HFQ) and therefore yields a tight bound on $x_\mathrm{qp}^\mathrm{array}$. We explicitly showed that accounting for a finite $\delta\Delta$ resolves this disparity in the case of \textit{injected} QPs. When $\delta\Delta \neq 0$, the suppression of the QP tunneling structure factor compensates for the strength of the junction array matrix element near the HFQ, making the inferred bound less tight. Whether a genuine difference exists between $x_\mathrm{qp}^\mathrm{array}$ and $x_\mathrm{qp}^\mathrm{small}$ for \textit{background} QPs remains an interesting open question. Unlike in the injection scenario, the background densities in the array and in the small junction are determined by the interplay of local QP generation and relaxation processes, which may differ between the two circuit elements.

We conclude by outlining several directions for future work. A readout scheme capable of distinguishing the leakage states would remove the main limitation of the present experiment and enable more reliable measurements of the QP-induced qubit excitation rates. This would provide better access to the energy dependence of the QP distribution function. It would also be interesting to study how the QP energy distribution and density change over time after injection. Furthermore, it remains to be explicitly demonstrated that SC gap engineering suppresses correlated errors produced by QP bursts in fluxonium, as it does in transmons~\cite{Connolly2024, McEwen2024, Nho2026}. Finally, QP-induced qubit frequency shifts~\cite{Catelani2011, Randeria2024} may affect the performance of quantum processors even with appreciable $\delta\Delta$~\cite{KurilovichV2026, Antonenko2026}. The effects of these shifts in fluxonium remain unexplored.

\textit{Note added.} Recently, we became aware of related theoretical work on QP-induced dissipation in fluxonium qubits~\cite{Azar2026Th}.

\section*{Acknowledgments}
We acknowledge helpful discussions with Kyle Serniak, Britton L.~T.~Plourde, Daniil S.~Antonenko, and Leonid I.~Glazman. This work is supported by the U.S. Government under ARO Grant No. W911NF-22-1-0257. This work was performed in part at the Cornell NanoScale Facility, a member of the National Nanotechnology Coordinated Infrastructure (NNCI), which is supported by the National Science Foundation (Grant NNCI-2025233).

\section*{Appendices}
\appendix

\section{Summary of previously reported bounds on quasiparticle densities in fluxonia}
\label{app:xqp_literature}

\begin{table*}[t]
\centering
\small
\begin{tabular}{cccccrrc}
\toprule
\shortstack{Device\\geometry} &
\shortstack{Substrate\\material} &
\shortstack{Ground\\plane} &
\shortstack{Junction\\style} &
\shortstack{Inductor\\type} &
$x_\mathrm{qp}^\mathrm{small}$\hspace{.6cm} &
$x_\mathrm{qp}^\mathrm{array/ind}$\hspace{.1cm} &
Ref.\\
\midrule

3D & sapphire & N/A & bridgeless & JJ array & $3.2\times10^{-6}{}^{*}$ & $-$\hspace{.9cm} & \cite{Pop2014} \\

3D & sapphire & N/A & bridgeless & JJ array & $-$\hspace{.9cm} & $4\times10^{-8}\phantom{{}^0}$ & \cite{Vool2014} \\

3D & sapphire & N/A & Dolan & grAl strip & $1.2\times10^{-5}{}^{*}$ & $4.1\times10^{-7}\phantom{{}^0}$ & \cite{Grunhaupt2019} \\

3D & Si & N/A & Dolan & JJ array & $-$\hspace{.9cm} & $1\times10^{-8}\phantom{{}^0}$ & \cite{Nguyen2019} \\

3D & sapphire & N/A & Dolan & JJ array & $5\times10^{-9}{}^{\dagger}$ & $6\times10^{-10}$ & \cite{Somoroff2023} \\

2D & sapphire & N/A & Dolan & grAl strip & $1.25\times10^{-6}\phantom{{}^0}$ & $-$\hspace{.9cm} & \cite{Atanasova2025} \\

2D & $-$ & $-$ & $-$ & $-$ & $4\times10^{-7}{}^{\dagger}$ & $5\times10^{-8}\phantom{{}^0}$ & \cite{Watanabe2025} \\

2D & Si & Al & Dolan & JJ array & $3\times10^{-6}\phantom{{}^0}$ & $7.4\times10^{-9}\phantom{{}^0}$ & \cite{Ateshian2025} \\

2D & Si & Al & Dolan & JJ array & $-$\hspace{.9cm} & $2.3\times10^{-8}{}^{*}$ & \cite{Zhuang2026} \\

2D & Si & Al & Dolan & JJ array & $1\times10^{-6}\phantom{{}^0}$ & $1\times10^{-8}\phantom{{}^0}$ & \cite{Azar2026Exp} \\

2D & sapphire & Al & Dolan & WSi nanowire & $-$\hspace{.9cm} & $3.9\times10^{-5}\phantom{{}^0}$ & \cite{Larson2026} \\

2D & Si & Nb & Dolan & JJ array & $4\times10^{-7}\phantom{{}^0}$ & $4\times10^{-8}{}^{*}$ & This work (Q1) \\

\bottomrule
\end{tabular}

\caption{
{\bf Quasiparticle-density bounds reported in fluxonium studies.}
The table summarizes bound estimates explicitly reported in the cited studies for the reduced QP densities in the small junction, $x_\mathrm{qp}^\mathrm{small}$, and in the inductor,
$x_\mathrm{qp}^\mathrm{array/ind}$, across different device implementations.
The ``3D'' label denotes devices with three-dimensional readout cavities, while ``2D'' denotes devices with on-chip readout resonators. The Josephson junctions in all cited works are of the Al/AlO$_\mathrm{x}$/Al type and are fabricated using either the Dolan-bridge~\cite{Dolan1977} or bridgeless~\cite{Lecocq2011} shadow-evaporation techniques. Inductors realized with Josephson junction (JJ) arrays are labeled as ``JJ array'', while those made of granular Al and WSi are labeled as ``grAl strip'' or ``WSi nanowire,'' respectively. Values marked with~``$^{*}$'' are upper bounds on $x_\mathrm{qp}$ inferred from $T_1$ while allowing for competing loss channels (method~2 in the text), whereas values marked with~``$^{\dagger}$'' are obtained from the $\ket{2}\!\to\!\ket{0}$ relaxation time, $T_{1}^{20}$ (method~3). All other values are obtained by assuming that QP loss is the dominant loss channel (method~1). ``N/A'' stands for not applicable, while ``$-$'' stands for no information available. The estimated upper bounds on the QP densities for qubit Q1 in this work are extracted from the background $T_1$ values under the conventional equal-gap assumption.}
\label{tab:qp_density_summary}
\end{table*}

Table~\ref{tab:qp_density_summary} summarizes the upper bound estimates of the reduced QP densities in the small junction $x_\mathrm{qp}^\mathrm{small}$ and in the inductor $x_\mathrm{qp}^\mathrm{array/ind}$ reported in the literature for fluxonium qubits. These studies examine QP effects at different levels of detail, ranging from studies focused specifically on QP-induced effects to those that consider QPs in the broader context of device performance. The bounds on $x_\mathrm{qp}$ can be divided into three main categories according to the method used to obtain them:
\begin{enumerate}
    \item An upper bound on $x_\mathrm{qp}$ is obtained by attributing all measured relaxation to QPs.
    \item An upper bound on $x_\mathrm{qp}$ is obtained from a model that includes QP loss together with competing loss channels.
    \item $x_\mathrm{qp}$ is inferred from $T_{1}^{20}$ at $\Phi_\mathrm{ext} = 0.5\,\Phi_0$ by attributing the measured $\ket{2}\!\to\!\ket{0}$ decay to QPs, since the symmetry leaves the relevant small-junction matrix element $\left|\mel{0}{\sin (\hat{\varphi}/2)}{2}\right|^2$ nonzero at the HFQ.
\end{enumerate}

In an early study, I.\,M.~Pop \textit{et al.}~\cite{Pop2014} measured the flux dependence of $T_1$ in a fluxonium qubit and observed an enhancement of $T_1$ near the HFQ, which was interpreted as a coherent suppression of QP tunneling across the small Josephson junction. The non-exponential decay of some $T_1$ curves was attributed to shot-to-shot fluctuations in the number of QPs. The bound on inductive loss $1/Q_\textrm{ind}$ reported in Ref.~\cite{Pop2014}, which the authors attributed mainly to QPs in the junction array, can be converted to a~bound on the QP density in the array $x_\mathrm{qp}^\mathrm{array} = 4\pi^3\sqrt{\hbar\omega_{01}/(2\Delta)}/Q_\mathrm{ind} \simeq 2.1\times10^{-8}$. Comparing this value with the corresponding bound for the small junction from the same work, $x_\mathrm{qp}^\mathrm{small} \simeq 3.2\times10^{-6}$, we find that the bound on the QP density in the small junction is almost two orders of magnitude larger than that in the junction array. In a subsequent experiment, U.\,Vool \textit{et al.}~\cite{Vool2014} continuously monitored a fluxonium qubit and observed alternating quiet and noisy periods in the quantum-jump traces, which were interpreted as temporal fluctuations of the QP population in the junction array. This work noted that the inferred bound on $x_\mathrm{qp}^\mathrm{array}$ was approximately an~order of magnitude lower than the bound on $x_\mathrm{qp}^\mathrm{small}$ reported in Ref.~\cite{Pop2014}; however, the origin of this difference remained unexplained.

Following these early experiments, several studies reported a~wide range of inferred bounds on QP densities across device implementations that differed in geometry, substrate, junction fabrication method, and superinductor material [Table~\ref{tab:qp_density_summary}]. Despite this variation, a recurring trend is that the upper bound on $x_\mathrm{qp}^\mathrm{small}$ is often \textit{at least an order of magnitude larger} than the corresponding bound on $x_\mathrm{qp}^\mathrm{array/ind}$. In this context, it is instructive to compare the fluxonium bounds on $x_\mathrm{qp}$ with estimates for transmon qubits. The reduced QP densities reported for transmon qubits also span several orders of magnitude, ranging from as low as $1.88\times10^{-11}$~\cite{Lin2026} and $5.6\times10^{-10}$~\cite{Connolly2024} to $1.3\times 10^{-8}$~\cite{Krause2024} and $5.16\times10^{-7}$~\cite{Pan2022}. We find that the bounds on $x_\mathrm{qp}^\mathrm{array/ind}$ and $x_\mathrm{qp}^\mathrm{small}$ in Table~\ref{tab:qp_density_summary} mostly fall within this range.

\section{General model of the quasiparticle-induced qubit transition rates}
\label{app:qp_rate_gen_model}

For completeness, we provide the derivation of Eq.~\eqref{eq:rates}, which forms the basis of the general model for QP-induced qubit transition rates used in the main text. We start by writing the expression for the QP-induced transition rates in the small junction~\cite{Catelani2012, Glazman2021, Connolly2024}:
\begin{equation}
\begin{split}
\Gamma_{if}^\mathrm{small} = 
\left|\mel{f}{\sin\frac{\hat{\varphi}}{2}}{i}\right|^2 \frac{E_J}h S^\mathrm{small}_+(\omega_{if})& \\
+ \left|\mel{f}{\cos\frac{\hat{\varphi}}{2}}{i}\right|^2 \frac{E_J}h S^\mathrm{small}_-(\omega_{if})&.
\end{split}
\label{eq:small_jj_rates}
\end{equation}
The QP tunneling structure factors $S^\alpha_{\pm}$, where generally $\alpha \in \{\mathrm{small},\mathrm{array}\}$, are expressed in terms of the BCS density of states $D(E,\Delta) = E\,\Theta(E-\Delta)/\sqrt{E^2-\Delta^2}$, where $\Theta(E)$ is the Heaviside step function, and the coherence factors:
\begin{equation}
\begin{split}
S^\alpha_{\pm}(\omega_{if}) = \frac{32}{\bar{\Delta}}
\int_0^{\infty} &dE\,
f^\alpha_L(E)\left[1-f^\alpha_H(E-\hbar\omega_{if})\right]  \\
&\times\,D(E,\Delta_L)\,D(E-\hbar\omega_{if},\Delta_H) \\
&\times\,F_{\pm}(E,E-\hbar\omega_{if}) \\
+ (L &\leftrightarrow H).
\end{split}
\label{eq:gen_model_via_dos}
\end{equation}
The coherence factors $F_{\pm}$ are given by
\begin{equation}
\begin{split}
F_{\pm}(E,E-\hbar\omega_{if}) = \frac {1}{2}
\left(1 \pm \frac{\Delta_L\Delta_H}{E(E-\hbar\omega_{if})}\right).
\end{split}
\end{equation}
Here, $E$ and $E - \hbar\omega_{if}$ are the initial and final QP energies, respectively, and $\omega_{if} = (E_f - E_i)/\hbar$ is the angular transition frequency between the qubit states $\ket{i}$ and $\ket{f}$. The functions $f_{L,H}(E)$ are the QP energy distributions in the low- and high-gap electrodes, while $\Delta_{L,H}$ denote the corresponding SC gaps. $\bar{\Delta} = (\Delta_L + \Delta_H)/2$ is the effective SC gap. The term $(L \leftrightarrow H)$ denotes the analogous contribution from QP tunneling from the high-gap to the low-gap electrode.

The expressions for the QP tunneling structure factors $S_{\pm}(\omega_{if})$ can be simplified under the following approximations: (i) $f_{L,H}(E) \ll 1$ and (ii) $\delta\Delta, \hbar\omega_{if}, \delta E_\mathrm{qp} \ll \bar{\Delta}$, where $\delta\Delta = \Delta_H - \Delta_L$ and $\delta E_\mathrm{qp}$ denotes the characteristic energy width of the QP distributions $f_{L,H}(E)$. The first approximation is justified by the low QP densities generated by injection in our experiments, $x_\mathrm{qp} \ll 1$. The second approximation is valid because the effective SC gap, $\bar{\Delta}/h \simeq 45\,\mathrm{GHz}$, is much larger than $\delta E_\mathrm{qp}/h$, $\delta\Delta/h$, and the relevant qubit transition frequencies.

Using the approximations introduced above and retaining only the leading-order terms in $\delta\Delta / \bar{\Delta}$, $\hbar\omega_{if} / \bar{\Delta}$, and $\delta E_\mathrm{qp} / \bar{\Delta}$, we obtain $1-f_{L,H}(E) \approx 1$, $F_{+} \approx 1$, and
$F_{-} \approx [-\hbar\omega_{if}-\delta\Delta+2(E-\Delta_L)]/(2\Delta_L)$.
Substituting these approximations into the expressions for the QP tunneling structure factors yields
\begin{equation}
\begin{split}
S^\alpha_+(\omega_{if}) \approx \frac{32}{\bar{\Delta}}
\int_0^{\infty} &dE\,
f^\alpha_L(E)\,D(E,\Delta_L)\,D(E-\hbar\omega_{if},\Delta_H) \\
+ (L &\leftrightarrow H), \\
S^\alpha_-(\omega_{if}) \approx \frac{32}{\bar{\Delta}}
\int_0^{\infty} &dE\,
f^\alpha_L(E)\,D(E,\Delta_L)\,D(E-\hbar\omega_{if},\Delta_H) \\
&\times \left(\frac{-\hbar\omega_{if} - \delta\Delta + 2(E-\Delta_L)}{2\Delta_{L}}\right)\\  
+ (L &\leftrightarrow H).
\end{split}
\label{eq:gen_model_via_dos_approx_final}
\end{equation}
Because the factor $\left[-\hbar\omega-\delta\Delta+2(E-\Delta_L)\right]/(2\Delta_L)$ entering $S^\alpha_{-}(\omega)$ is parametrically small, we have $S^\alpha_{-}(\omega) = \mathcal{O}\!\left(\frac{|\hbar\omega|}{\bar{\Delta}}, \frac{\delta\Delta}{\bar{\Delta}}, \frac{\delta E_\mathrm{qp}}{\bar{\Delta}}\right) S^\alpha_{+}(\omega)$, and therefore $S^\alpha_{-}(\omega) \ll S^\alpha_{+}(\omega)$.

The junction array contribution is obtained by summing the contributions from all $N$ array junctions and accounting for the phase drop $\hat{\varphi}_\mathrm{array}^{(1)}$ across an individual junction in the array, $\hat{\varphi}_\mathrm{array}^{(1)} \approx \hat{\varphi}_\mathrm{array}/N = (\hat{\varphi}-\varphi_\mathrm{ext})/N$. Replacing $\hat{\varphi}$ by $\hat{\varphi}_\mathrm{array}^{(1)}$ in Eq.~\eqref{eq:small_jj_rates}, we obtain
\begin{equation}
\begin{split}
\Gamma_{if}^\mathrm{array} \approx 
N \times \Bigg[
\left|\mel{f}{\sin\frac{\hat{\varphi}-\varphi_\mathrm{ext}}{2N}}{i}\right|^2 \frac{E_J^\mathrm{array}}h S^\mathrm{array}_+(\omega_{if})& \\
+ \left|\mel{f}{\cos\frac{\hat{\varphi}-\varphi_\mathrm{ext}}{2N}}{i}\right|^2 \frac{E_J^\mathrm{array}}h S^\mathrm{array}_-(\omega_{if})&
\Bigg].
\end{split}
\label{eq:array_jj_rates}
\end{equation}
Keeping only the leading-order terms in the expansions of the sine and cosine and using the orthogonality of the eigenstates $\ip{f}{i} = \delta_{if}$, we obtain the final expression for the junction array contribution to the QP-induced transition rates~\cite{Catelani2011}:
\begin{equation}
\begin{split}
\Gamma_{if}^\mathrm{array} \approx \frac{1}{N}\left|\mel{f}{\frac{\hat{\varphi}}{2}}{i}\right|^2 \frac{E_J^\mathrm{array}}{h} S^\mathrm{array}_+(\omega_{if}).
\end{split}
\label{eq:array_jj_rates_approx}
\end{equation}
Finally, the factor of $1/N$ is absorbed into the definition $E_L = E_J^\mathrm{array}/N$, yielding the junction array contribution in the form used in Eq.~\eqref{eq:rates}.

\section{Device fabrication}
\label{app:fab}
The qubit devices are fabricated on 100-mm-diameter undoped high-resistivity (100) silicon wafers with a standard thickness of $525\pm10 \,\mathrm{\mu m}$, following the fabrication steps described below.

\textit{Wafer pre-cleaning.} The wafers are submerged in a 5\% hydrofluoric acid (HF) solution for 1\,min, rinsed with deionized (DI) water for 1\,min, gun-dried with nitrogen gas, and immediately loaded into a sputtering system to minimize surface reoxidation.

\textit{Nb sputtering.} After loading the wafer into the AJA Orion sputtering system, the sputtering chamber is pumped out overnight to reach a base pressure in the low $10^{-8}$\,Torr range. Prior to deposition, pre-sputtering is performed to clean the niobium target and the chamber interior. Following this pre-conditioning step, approximately 75\,nm of niobium is deposited on the wafer surface.

\textit{Photolithography.} After removal from the sputtering system, a layer of anti-reflective coating (ARC) is spun onto the wafer and baked on a hot plate to prevent overexposure during lithography. After this bake, a layer of DUV210 deep-ultraviolet (DUV) resist is deposited, spun, and cured on a hot plate. The wafer is then loaded into the ASML DUV lithography system (stepper), where all Nb-layer features are patterned. Following exposure, the wafer undergoes a post-exposure bake and is subsequently developed in 726MIF developer solution, removing the exposed DUV210 resist. The wafer is inspected and imaged after development to verify that the patterned features are well-defined prior to etching.

\textit{Nb etching.} Prior to ARC removal, the chamber of the Oxford Cobra inductively coupled plasma reactive ion etching (ICP RIE) system is prepared with a 10-min oxygen plasma clean to minimize residual contamination from previous use. The ARC is then etched from the device wafer. In parallel, a Nb-coated dummy wafer is used to pre-condition a separate PT770 Plasma-Therm SLR etcher. After ARC removal from the device wafer and sputter pre-cleaning of the PT770 chamber, the device wafer is transferred as quickly as possible to the PT770 etcher, where the photolithographic patterns are etched into the niobium layer.

\textit{Intermediate cleaning.} Prior to electron-beam lithography, the wafer is stripped of remaining resist in a tetramethylammonium hydroxide (TMAH) hot-strip bath. The wafer is immersed in a first TMAH bath for 15\,min, transferred to a second TMAH bath for an additional 15\,min, and then rinsed in a DI water dump-rinse tank until the wastewater reaches an acceptable resistivity, indicating sufficient cleanliness. The wafer is then dried and loaded into a Glenn 1000-F Series etcher for a~light descum to remove residual resist.

\textit{Electron-beam lithography.} After complete resist removal, the wafer undergoes an additional cleaning step consisting of a 1-min dip in 5\% HF followed by a 1-min rinse in a DI water bath. The wafer is dried with a nitrogen gun and immediately coated with a methyl methacrylate (MMA) layer, which is spun and cured on a hot plate. A poly(methyl methacrylate) (PMMA) layer is then deposited, spun, and cured in the same manner. After preparation of the MMA/PMMA bilayer resist stack, the wafer is loaded into a JEOL 6300 electron-beam lithography (EBL) tool. Prior to exposure, the electron beam is tuned to a spot size below 20\,nm.

\textit{Double-angle Al evaporation.} To target specific device parameters, the wafer is first segmented into several pieces using a diamond scribe, enabling test evaporation runs prior to the final evaporation. Each wafer segment is developed in a 4:1 deionized water:isopropyl alcohol (DI:IPA) bath held at $2\,^\circ\mathrm{C}$ in an ice bath for approximately 2\,min. The wafer segment is then dried with a~nitrogen gun and transferred to the evaporation chamber as quickly as possible. The chamber is pumped overnight, after which the wafer segment is ion-milled to remove oxides that may have formed near the junction regions during processing. The aluminum crucible is then heated by an electron beam until a stable aluminum deposition rate is reached, as measured by a film-thickness monitor. Once the deposition rate stabilizes at approximately $2 \,\mathring{\mathrm{A}}/\mathrm{s}$, the shutter is opened, and 40\,nm of aluminum is deposited at the desired angle. The electron beam is then turned off, and the chamber is filled with a 90\% argon and 10\% oxygen mixture to form the oxide tunnel barrier of the Josephson junctions. For a given run, the pressure of the gas mixture is nominally fixed within the 10--100\,Torr range, while the oxidation time, typically 10--180\,min, is adjusted to achieve the target critical current density. The chamber is then evacuated, the wafer segment is rotated to the second angle, and the final 80\,nm aluminum layer is deposited in the same manner as the first.

\textit{Liftoff.} A layer of S1813 resist is spun onto the wafer and lightly baked to protect the devices from dicing debris. With this protective layer in place, the wafer segment is diced into individual chips. Liftoff is then performed in a two-stage Remover PG bath to ensure removal of the remaining S1813 resist while minimizing sample contamination. This cleaning step is carried out on a hot plate set to $60\,^\circ\mathrm{C}$ for up to 12~hours. The sample is then rinsed and sonicated, first in IPA and then in DI water, for 1\,min each.

\textit{Sample packaging.} Prior to packaging, each sample is visually inspected for large-scale defects and then mounted to the base of an aluminum sample box by applying GE varnish to the chip corners. After the GE varnish cures at room temperature, aluminum wire bonds are made between the copper packaging leads and the on-chip signal lines, while the ground plane is uniformly wire-bonded along the perimeter. A light-tight lid is then screwed onto the top of each box before the sample is loaded into the dilution refrigerator for measurements.

\section{Experimental setup}
\label{app:exp_setup}

\begin{figure*}
 \centering
 \includegraphics[width=0.90\textwidth]{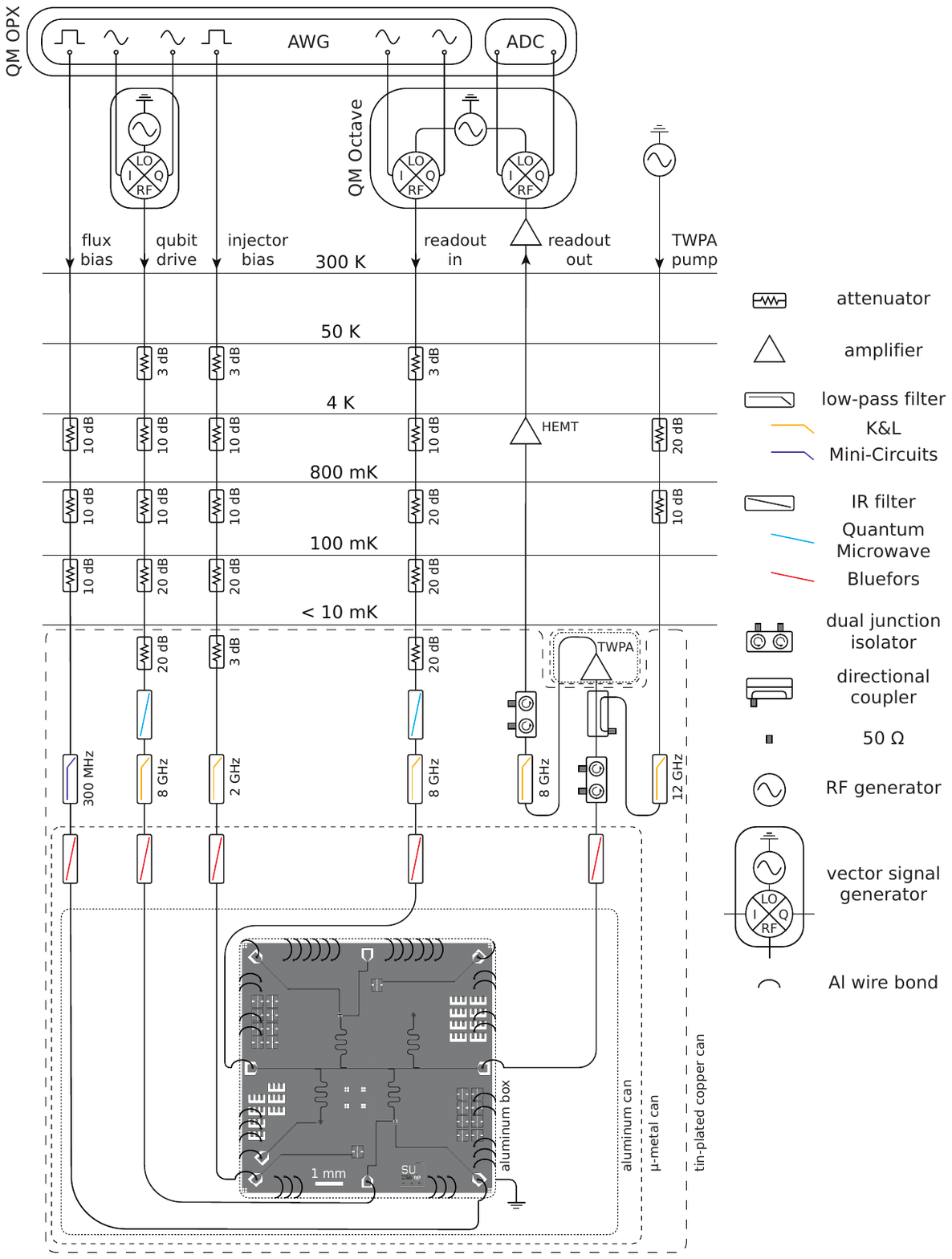}
 \caption{{\bf Wiring diagram of the dilution refrigerator measurement setup.} The chip is enclosed in an aluminum sample box. In the false-colored chip image, the gray regions indicate the Nb ground plane, while the white regions correspond to exposed Si. Unused control lines on the chip are shorted to ground with ground caps. Magnetic and radiation shields are indicated by distinct line styles.}
 \label{fig:exp_setup}
\end{figure*}

The experiments are performed in a Bluefors LD400 dilution refrigerator (DR) at temperatures below $10 \,\mathrm{mK}$. Figure~\ref{fig:exp_setup} shows a schematic of the room-temperature (RT) control electronics and the wiring configuration inside the DR. The samples are packaged in aluminum sample boxes, as described in Appendix~\ref{app:fab}, and mounted on the copper cold finger, which is thermally anchored to the mixing chamber (MXC) stage of the DR.

The baseband input signals for the qubit drive and readout are generated using the Quantum Machines (QM) OPX arbitrary waveform generator (AWG). The readout tone is upconverted with the QM Octave, and the qubit drive tone is upconverted using a Rohde \& Schwarz SGS100A SGMA RF source configured as a vector signal generator. Bias voltages for the flux bias line and injector junction are applied directly using the OPX analog outputs.

Inside the DR, all control and measurement lines are equipped with cryogenic XMA attenuators distributed across different temperature stages, K\&L Microwave and Mini-Circuits low-pass filters (LPFs), and Bluefors and Quantum Microwave ECCOSORB infrared (IR) filters mounted at the MXC stage. The attenuators and filters at the MXC stage are thermally anchored to the plate using oxygen-free high-conductivity (OFHC) copper clamps. Magnetic shielding is provided by a Bleximo $\mu$-metal can and an aluminum superconducting shield, both thermally anchored to the MXC plate with silver-coated copper braids. The inner surface of the aluminum shield is coated with an IR-absorbing material to suppress stray IR radiation. The entire shield assembly is enclosed in a tin-plated copper can.

The readout signal exiting the sample package passes through an ECCOSORB IR filter and an LNF-ISISC4\_8A dual-junction Low Noise Factory (LNF) isolator before being routed to the traveling-wave parametric amplifier (TWPA) provided by MIT Lincoln Laboratory. The TWPA is enclosed in a separate Bleximo $\mu$-metal can and thermally anchored to the MXC stage. A~QMC-CRYOCOUPLER-20 Quantum Microwave directional coupler combines the TWPA pump with the readout tone. The pump tone is generated by a~DS Instruments SG22000PRO signal generator and injected into the isolated port of the coupler, which provides an additional $20~\mathrm{dB}$ of attenuation before reaching the TWPA. After TWPA amplification, the signal is filtered using a K\&L LPF, passed through a second LNF-ISISC4\_8A dual-junction isolator, and then amplified by an LNF-LNC4\_8C high-electron-mobility transistor (HEMT) amplifier at the $4 \,\mathrm{K}$ stage. Further amplification is performed at room temperature using a Mini-Circuits ZX60-83LN-S+ low-noise amplifier.

The amplified readout signal is downconverted using the Octave, and the resulting $I(t)$ and $Q(t)$ quadratures are digitized by the OPX analog-to-digital converter (ADC). The signal is then digitally demodulated to obtain the readout quadrature components $I$ and $Q$. The signal $V_\mathrm{ADC}=I+iQ$ is proportional to the $S_{21}$ parameter and encodes the qubit state via the state-dependent dispersive shift of the readout resonator.

Device Q1, discussed in the main text, was measured using the experimental setup described above. Devices Q2--Q4, presented in Appendix~\ref{app:qubits}, were measured with only minor variations in the wiring configuration and electronics.

\section{Qubits}
\label{app:qubits}

\subsection{Fluxonia}

\begin{figure*}
 \centering
 \includegraphics[width=\textwidth]{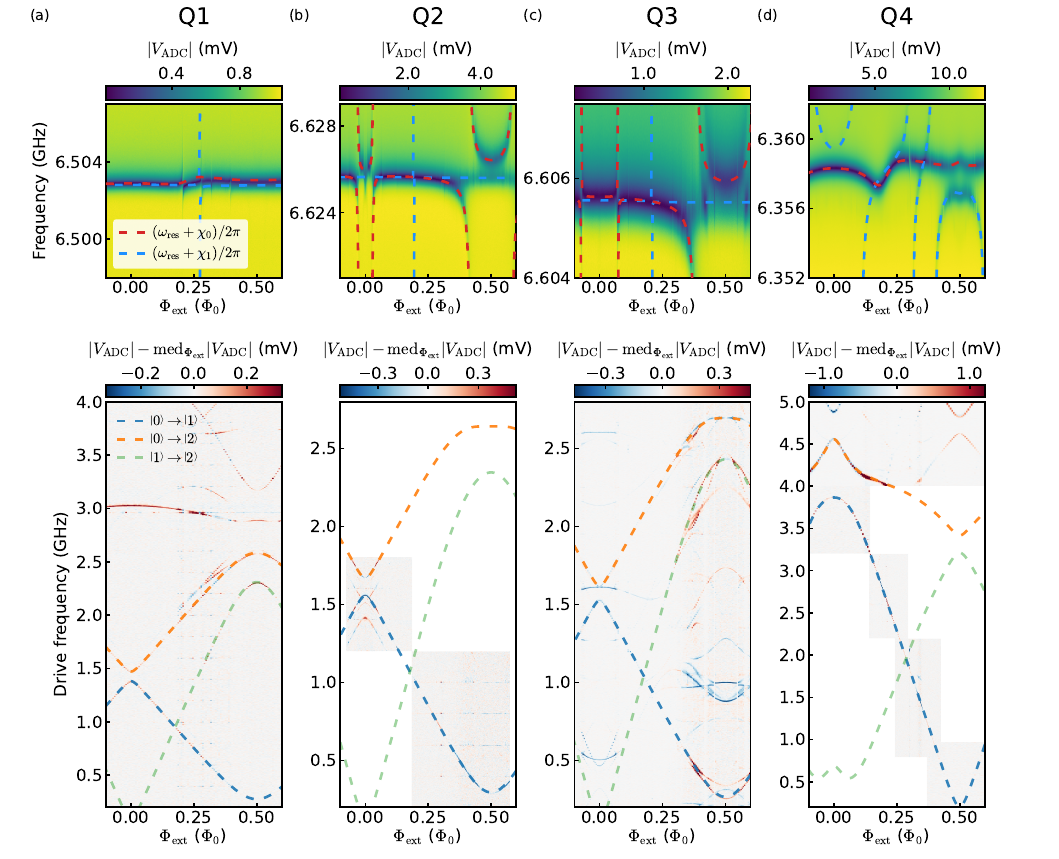}
 \caption{\textbf{Resonator and qubit spectroscopy of fluxonium qubits Q1--Q4.} (a)--(d), top panels: One-tone spectroscopy of the readout resonators as a function of probe frequency and external flux $\Phi_\mathrm{ext}$. The red and blue dashed curves indicate the fitted qubit-state-dependent resonator frequencies for the qubit in the $\ket{0}$ and $\ket{1}$ states, respectively. (a)--(d), bottom panels: Two-tone qubit spectroscopy as a function of qubit drive frequency and external flux $\Phi_\mathrm{ext}$. At each value of $\Phi_\mathrm{ext}$, the median value of $|V_{\mathrm{ADC}}|$ is subtracted from the data at that flux to improve the visibility of the spectroscopic features in the two-dimensional plots. The dashed curves show the fitted fluxonium transition frequencies for the $\ket{0}\!\to\!\ket{1}$ (blue), $\ket{0}\!\to\!\ket{2}$ (orange), and $\ket{1}\!\to\!\ket{2}$ (green) transitions. The curves are obtained by fitting the spectra with the coupled-system Hamiltonian~\eqref{eq:H_system}.}
 \label{fig:spectra}
\end{figure*}

\begin{table}[t]
\centering
\small
\renewcommand{\arraystretch}{1.2}
\begin{tabular}{ccccc}
\toprule
Parameter & Q1 & Q2 & Q3 & Q4 \\
\midrule

$E_J/h$ (GHz) & 2.043 & 1.994 & 2.124 & 3.172 \\
$E_C/h$ (GHz) & 1.086 & 1.083 & 1.071 & 0.962 \\
$E_L/h$ (GHz) & 0.078 & 0.089 & 0.086 & 0.269 \\
$\omega_\mathrm{res}/2\pi$ (GHz) & 6.503 & 6.626 & 6.605 & 6.357 \\
$g/2\pi$ (MHz) & 63 & 64 & 57 & 137 \\
$\kappa/2\pi$ (MHz) & 0.8 & 0.8 & 0.8 & 1.3 \\

\midrule

$w_\mathrm{small}$ ($\mathrm{\mu m}$) & 0.17 & 0.17 & 0.17 & 0.18 \\
$l_\mathrm{small}$ ($\mathrm{\mu m}$) & 0.40 & 0.40 & 0.40 & 0.10 \\
$w_\mathrm{array}$ ($\mathrm{\mu m}$) & 1.10 & 1.10 & 1.10 & 1.50 \\
$l_\mathrm{array}$ ($\mathrm{\mu m}$) & 0.40 & 0.40 & 0.40 & 0.14 \\
$N$ & 221 & 221 & 221 & 203\\

\bottomrule
\end{tabular}

\caption{\textbf{Qubit parameters.}
Device parameters for the four fluxonium qubits Q1--Q4 studied in this work. The extracted fluxonium parameters, $E_J/h$, $E_C/h$, and $E_L/h$, the qubit--resonator coupling strength $g/2\pi$, and the bare resonator frequency $\omega_\mathrm{res}/2\pi$ are obtained from a~fit of the spectroscopic lines to the transitions of a coupled qubit--resonator system. The resonator linewidth $\kappa/2\pi$ is extracted from a circle fit to the complex resonator transmission response. $w_\mathrm{small}$ and $w_\mathrm{array}$ are the small- and array-junction widths, respectively, whereas $l_\mathrm{small}$ and $l_\mathrm{array}$ are the corresponding junction lengths defined by double-angle deposition. $N$ denotes the number of Josephson junctions in the junction array.}
\label{tab:qubit_parameters}
\end{table}

Table~\ref{tab:qubit_parameters} summarizes the parameters of the four fluxonium devices from four different chips studied in this work. The qubit and resonator parameters are obtained from a fit to the qubit and resonator spectra shown in Fig.~\ref{fig:spectra}, using the coupled system Hamiltonian~\cite{Zhu2013}:
\begin{equation}
\hat{H}_\mathrm{system}=
\hat{H}_\mathrm{qubit}+\hat{H}_\mathrm{res}+\hat{H}_\mathrm{int}.
\label{eq:H_system}
\end{equation}
Here, $\hat{H}_\mathrm{qubit}$ is the qubit Hamiltonian defined in Eq.~\eqref{eq:hamiltonian}. The resonator Hamiltonian is
$\hat{H}_\mathrm{res}=\hbar\omega_\mathrm{res}\left(\hat{a}^\dagger\hat{a}+1/2\right)$, where $\omega_\mathrm{res}$ is the bare angular frequency of the resonator and $\hat{a}$ and $\hat{a}^\dagger$ are the annihilation and creation operators, respectively, satisfying $[\hat{a}, \hat{a}^\dagger]=1$. The capacitive coupling between the qubit and the resonator is described by $\hat{H}_\mathrm{int}=-i\hbar g\hat{n}(\hat{a}-\hat{a}^\dagger)$, where $g$ is the qubit-resonator coupling strength.

In addition, Table~\ref{tab:qubit_parameters} lists the geometric dimensions of the qubit junctions. The small- and array-junction widths, $w_\mathrm{small}$ and $w_\mathrm{array}$, respectively, are nominal values taken from the design layout. The corresponding junction lengths $l_\mathrm{small}$ and $l_\mathrm{array}$ are due to the overlap of the leads produced by double-angle deposition. These lengths are extracted from scanning electron micrographs of test junctions fabricated in the same run.

For all qubits, the junction-lead thicknesses are 40\,nm for the bottom lead and 80\,nm for the top lead, as determined from the film-thickness monitor during double-angle evaporation. To estimate the SC gap asymmetry $\delta\Delta$ between the two thin-film Al junction leads, we use the relation $\Delta(t) = \Delta_\mathrm{bulk} + a/t$, w here $\Delta_\mathrm{bulk} \approx 180 \,\mu\mathrm{eV}$ is the bulk Al gap and $a \approx 600\,\mu\mathrm{eV}{\cdot}\mathrm{nm}$~\cite{Marchegiani2022}. We obtain $\delta\Delta/h \approx 1.8\,\mathrm{GHz}$ for the Josephson junctions studied in this work.

\subsection{Transmons}
\label{app:qubits_and_transmons}

All four chips with fluxonium qubits Q1--Q4 also include transmons~\cite{Koch2007}. For brevity, we refer only to the transmon co-fabricated with qubit Q1 on the same chip, which was used to estimate the effective temperature of the QP energy distribution [Sec.~\ref{sec:qp_energy_distribution}]. The transmon parameters are $E_J/h = 7.278\,\mathrm{GHz}$ and $E_C/h = 0.293\,\mathrm{GHz}$.

\section{Photon-mediated quasiparticle poisoning}
\label{app:photon_poisoning}

\begin{figure}
 \includegraphics[width=\columnwidth]{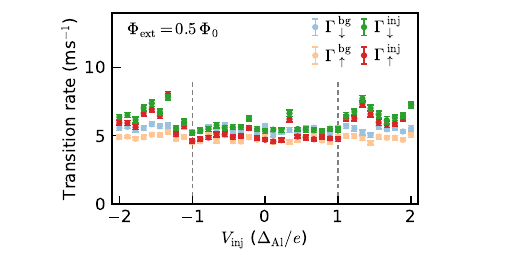}
 \caption{\textbf{Josephson-radiation-induced transition rates in qubit Q1.} Background transition rates, $\Gamma_\downarrow^\mathrm{bg}$ (light blue) and $\Gamma_\uparrow^\mathrm{bg}$ (light orange), and transition rates during injection, $\Gamma_\downarrow^\mathrm{inj}$ (green) and $\Gamma_\uparrow^\mathrm{inj}$ (red), as functions of injection bias $V_\mathrm{inj}$ at the HFQ. To resolve the injection-induced response when the injection bias falls within the doubled SC gap, $|V_\mathrm{inj}| \leq 2\Delta_\mathrm{Al}/e$, a long saturating injection of duration $t_\mathrm{inj} = 1.0 \, \mathrm{ms}$ is applied before qubit state preparation and kept on until the final readout. For $|V_\mathrm{inj}| \leq \Delta_\mathrm{Al}/e$, the injection transition rates remain nearly flat and coincide with the background transition rates. Near $V_\mathrm{inj} = \pm\Delta_\mathrm{Al}/e$, the transition rates exhibit a clear onset of Josephson-radiation-induced transitions. For $\Delta_\mathrm{Al}/e < |V_\mathrm{inj}| < 2\Delta_\mathrm{Al}/e$, the transition rates show a nonlinear response with a peaked structure characteristic of resonant absorption of coherent Josephson radiation. The transition rates are extracted from a joint fit to the population data in a wide $0 \leq t_\mathrm{delay} \leq 540 \,\mathrm{\mu s}$ time window, as no time dependence of the rates is expected in the injection-saturated regime. Gray dashed lines mark $V_\mathrm{inj} = \pm\Delta_\mathrm{Al}/e$.}
 \label{fig:jj_radiation}
\end{figure}

In addition to phonon-mediated QP poisoning, the injector junction can emit CP-breaking photons, leading to photon-mediated QP poisoning. In this section, we discuss the two types of radiation generated by the voltage-biased injector junction: coherent Josephson radiation and incoherent broadband photon emission~\cite{Liu2024}. Both can generate QPs at the qubit either through photon-assisted tunneling across the junctions~\cite{Houzet2019} or through direct CP-breaking in the SC films~\cite{Guruswamy2015, Fischer2024}.

For a Josephson junction dc-biased at a voltage $V_\mathrm{inj}$, the SC phase difference between the leads evolves according to the ac Josephson relation $\dot{\varphi} = 2eV_\mathrm{inj}/\hbar$. The resulting ac supercurrent $I_s(t) = I_c\sin(2\pi f_J t + \phi_0)$ oscillates with the Josephson frequency $f_J = 2eV_\mathrm{inj}/h$, where $I_c$ is the critical current of the junction. Microscopically, this emission can be viewed as inelastic CP tunneling across the voltage-biased junction, in which the energy $2e|V_\mathrm{inj}|$ released by each tunneling CP is transferred to the electromagnetic environment as photons~\cite{Holst1994, Cassidy2017}.

It should be noted that applying an injector voltage of $|V_\mathrm{inj}| = \Delta_\mathrm{Al}/e$, which is below the threshold for generating CP-breaking \textit{phonons}, results in $hf_J = 2\Delta_\mathrm{Al}$ and therefore allows the emission of CP-breaking \textit{photons}. To verify this, we perform a long saturation injection with $t_\mathrm{inj} = 1.0 \,\mathrm{ms}$ and $|V_\mathrm{inj}| < 2\Delta_\mathrm{Al}/e$ before the beginning of our measurement sequence [Fig.~\ref{fig:protocol}(a)] and keep the injector on throughout the sequence, including during the final readout. Figure~\ref{fig:jj_radiation} shows the measured background transition rates and transition rates during injection. Consistent with expectations, we observe $\Gamma_{\downarrow\,,\uparrow}^\mathrm{inj}\,>\, \Gamma_{\downarrow\,,\uparrow}^\mathrm{bg}$ starting at $|V_\mathrm{inj}|=\Delta_\mathrm{Al}/e$. However, the resulting photon-mediated QP poisoning at $|V_\mathrm{inj}|<2\Delta_\mathrm{Al}/e$ is weak [Fig.~\ref{fig:rates_inj_bias}(b,\,c)] and becomes detectable only for injection durations much longer than those used elsewhere in this work.

Another feature of coherent Josephson radiation is that high-energy photons emitted by the injector junction, which acts as a transmitter, can be resonantly absorbed by spurious antenna modes in the qubit circuit, which acts as a receiver, thereby efficiently generating QPs. Qubit antenna modes and the associated resonant photon-mediated QP-poisoning mechanism have been studied in detail in transmon qubits, both theoretically in Ref.~\cite{Rafferty2021} and experimentally in Refs.~\cite{Pan2022, Liu2024}. In many of the devices studied, we observe weak peaks in the post-injection transition rates, which we associate with resonant photon absorption [Figs.~\ref{fig:rates_inj_bias}(b,\,c) and~\ref{fig:jj_radiation}]. The most pronounced peak attributed to the qubit–injector antenna mode was observed in device Q3 and, to maximize the detectable response, we set $V_\mathrm{inj}$ to the voltage corresponding to this peak when investigating the external flux dependence of the QP-induced transition rates in this device [Appendix~\ref{app:simple_model}, Fig.~\ref{fig:simple_xqp_fits}(c) and Appendix~\ref{app:asymmetric_gap_model}, Fig.~\ref{fig:xqp_fits}(c)].

Besides coherent Josephson radiation, shot noise associated with QP tunneling across the injector junction can produce incoherent broadband photon emission for $|V_\mathrm{inj}| \ge 2\Delta_\mathrm{Al}/e$, with the maximum photon energy given by $e|V_\mathrm{inj}| - 2\Delta_\mathrm{Al}$. For $|V_\mathrm{inj}| \ge 4\Delta_\mathrm{Al}/e$, the shot-noise emission spectrum contains CP-breaking photons, and this incoherent contribution can dominate over coherent Josephson radiation with increasing injection bias~\cite{Liu2024}. As in the case of coherent Josephson radiation, however, we do not observe a pronounced effect of incoherent broadband radiation in the injection experiments.

\section{Post-injection recovery}
\label{app:t1_recovery}

\begin{figure}
 \centering
 \includegraphics[width=\columnwidth]{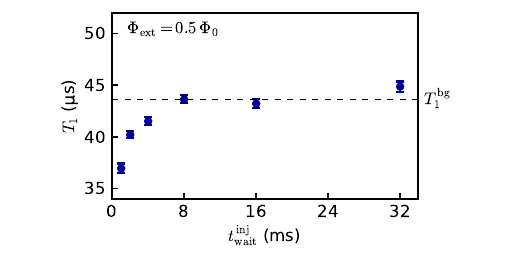}
 \caption{\textbf{Post-injection recovery of the energy relaxation time $T_1$.} Apparent $T_1$ of qubit Q2 measured at $\Phi_\mathrm{ext} = 0.5\,\Phi_0$ as a~function of the post-injection wait time $t_\mathrm{wait}^\mathrm{inj}$ in the protocol shown in Fig.~\ref{fig:protocol}(b). At short wait times, $T_1$ is suppressed by residual excess QPs but gradually recovers toward the background value $T_1^\mathrm{bg}$, which is measured in a~separate experiment in the absence of QP injection in the protocol. Injection parameters: $V_\mathrm{inj} = 13.8\,\Delta_\mathrm{Al}/e$, $t_\mathrm{inj} = 1.0\,\mu\mathrm{s}$.}
 \label{fig:t1_vs_inj_wait_time}
\end{figure}

The excess QPs and athermal phonons in the substrate generated by injection relax to their background values over a~range of timescales that depend on the device materials, geometry, and thermal anchoring~\cite{Wang2014, Wilen2021, McEwen2022, Iaia2022, Harrington2025, Pinckney2026, Yang2026}. Therefore, an appropriate post-injection wait time $t_\mathrm{wait}^\mathrm{inj}$ must be chosen to ensure that the QP density has returned to its baseline level and no longer affects subsequent measurements.

Figure~\ref{fig:t1_vs_inj_wait_time} shows $T_1$ as a function of the post-injection wait time $t_\mathrm{wait}^\mathrm{inj}$, as defined in Fig.~\ref{fig:protocol}(b). For short $t_\mathrm{wait}^\mathrm{inj}$ on the order of $1\,\mathrm{ms}$, the effect of injection on the inferred $T_1$ remains clearly visible, indicating the presence of residual effects caused by injection. For $t_\mathrm{wait}^\mathrm{inj} \gtrsim 8\,\mathrm{ms}$, $T_1$ approaches the expected background value $T_1^\mathrm{bg}$, which is obtained in a~measurement without injection and therefore faithfully represents the background energy relaxation time. Based on these observations, we chose $t_\mathrm{wait}^\mathrm{inj} \ge 16\,\mathrm{ms}$ in our measurements to ensure that QP densities had recovered to their background levels.

We note that our post-injection recovery timescale is compatible with that inferred from previous measurements in transmon qubits~\cite{Wang2014}. The recovery can be due to QP recombination or trapping, and the observed timescale indicates that the latter mechanism is likely responsible for the recovery. Indeed, the timescale associated with QP recombination can be estimated as $1/(r x_\mathrm{qp})$, where the recombination coefficient in Al is of order $r^{-1} \sim 10$--$100\,\mathrm{ns}$. For an injected QP density $x_\mathrm{qp} \sim 10^{-6}$, this suggests a recombination time much longer than $10\,\mathrm{ms}$, since recombination slows down as the QP density decreases. By contrast, residual trapping (that is, trapping in the absence of vortices) has been reported to occur on a timescale $s^{-1}\sim 10\,\mathrm{ms}$, which is of the same order as the recovery time in our experiment.

\section{Simplified model}
\label{app:simple_model}

\begin{figure*}
 \centering
 \includegraphics[width=\textwidth]{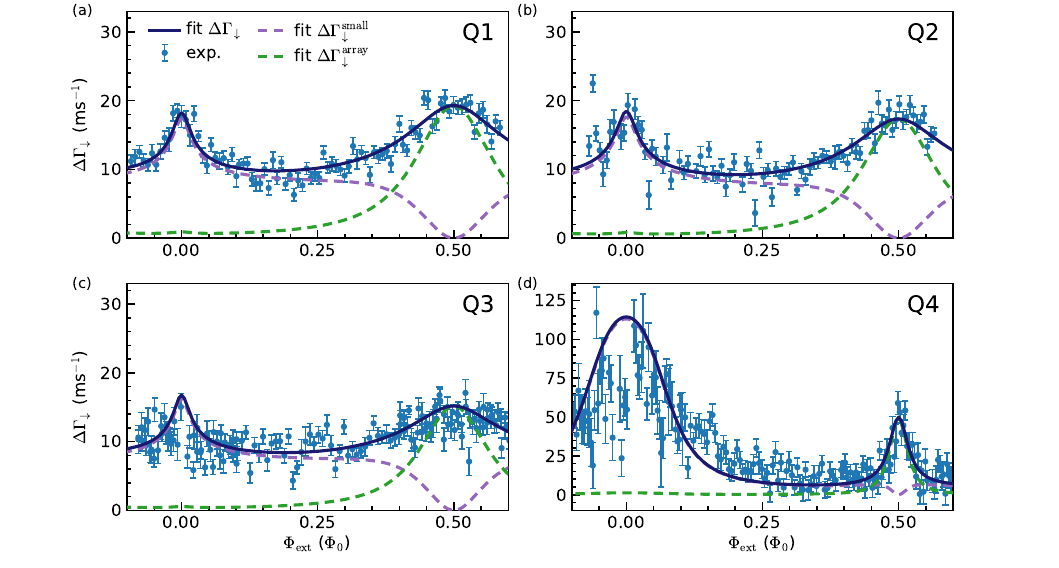}
 \caption{\textbf{Fits of the quasiparticle-induced de-excitation rates $\Delta\Gamma_\downarrow$ to a simplified model for qubits Q1--Q4.} (a)--(d) QP-induced de-excitation rates ($\Delta\Gamma_\downarrow = \Gamma_\downarrow^\mathrm{inj} - \Gamma_\downarrow^\mathrm{bg}$) as a function of external flux~$\Phi_\mathrm{ext}$ for Q1--Q4. The blue solid lines show least-squares fits of the simplified model to the experimental $\Delta\Gamma_\downarrow$ data. The purple and green dashed lines denote the contributions in the fits from the small junction and junction array, respectively. The injection parameters are the same as in Fig.~\ref{fig:xqp_fits}. The y-axis in panel (d) is adjusted to show the full range of measured de-excitation rates for Q4.}
 \label{fig:simple_xqp_fits}
\end{figure*}

The works summarized in Table~\ref{tab:qp_density_summary} infer bounds on the QP densities in the small junction and in the junction array, using QP-induced transition rate models based on simplified approximations. Here, we list these assumptions, present a particular expression for the rate $\Gamma_{10}^\mathrm{simple}$, and apply this expression to fit our experimental data to extract the QP densities in the small junction and the junction array within the same simplified framework.

The simplified model relies on the following assumptions~\cite{Catelani2011, Glazman2021}: (i)~the SC gaps on the two sides of the junction are equal, i.e., $\Delta_L = \Delta_H \equiv \Delta$; (ii)~the qubit transition energy $\hbar\omega_{01}$ is small compared to the SC gap, i.e., $\hbar\omega_{01} \ll \Delta$; (iii)~the qubit transition energy $\hbar\omega_{01}$ is larger than the width of the QP energy distribution $\delta E_\mathrm{qp}$, i.e., $\delta E_\mathrm{qp} \ll \hbar\omega_{01}$.

The third assumption allows us to consider nonthermal QP distributions and characterize them by the total reduced QP density $x^\alpha_\mathrm{qp}$, $\alpha \in \{\mathrm{small},\mathrm{array}\}$. In this limit, the QP tunneling structure factors are given by $S_+^{\alpha}(\omega_{10}) = 16x_\mathrm{qp}^{\alpha}\sqrt{2\Delta/(\hbar\omega_{01})}$ and $S_+^{\alpha}(\omega_{01}) \approx 0$, reflecting the strong suppression of qubit excitation due to QPs. Using Eq.~\eqref{eq:rates}, the resulting total QP-induced de-excitation rate is
\begin{equation}
\label{eq:simple_rates}
\begin{split}
\Gamma_{10}^\mathrm{simple}
= \left|\mel{0}{\sin\frac{\hat{\varphi}}{2}}{1}\right|^2 \frac{16E_{J}}{h} x_\mathrm{qp}^\mathrm{small} \sqrt{\frac{2\Delta}{\hbar\omega_{01}}}&  \\ 
+ \left|\mel{0}{\frac{\hat{\varphi}}{2}}{1}\right|^2 \frac{16E_{L}}{h} x_\mathrm{qp}^\mathrm{array} \sqrt{\frac{2\Delta}{\hbar\omega_{01}}}&,
\end{split}
\end{equation}
where the first and second terms correspond to the contributions from the small junction and the junction array, respectively. These separate contributions are commonly used to place upper bounds on the corresponding QP densities in fluxonium qubits~\cite{Pop2014, Vool2014, Grunhaupt2019, Nguyen2019, Somoroff2023, Watanabe2025, Azar2026Exp, Larson2026}. However, some works replace assumption~(iii) with the weaker condition $\delta E_\mathrm{qp} \ll \Delta$ and, assuming a thermal QP distribution, introduce an effective QP-bath temperature that is then used to estimate the QP densities in the small junction and the junction array~\cite{Atanasova2025, Ateshian2025}.

To make a direct comparison with previous reports, we fit the simplified model of QP-induced transition rates given by Eq.~\eqref{eq:simple_rates} to our data, as shown in Fig.~\ref{fig:simple_xqp_fits}, using $x_\mathrm{qp}^\mathrm{small}$ and $x_\mathrm{qp}^\mathrm{array}$ as fitting parameters. We only fit QP-induced de-excitation rates $\Delta\Gamma_\downarrow$ because within this simplified model the excitation rates are strongly suppressed, i.e., $\Gamma^\mathrm{simple}_{01} \approx 0$. The fits converge for all qubits studied in this work and follow a characteristic two-peak profile, arising from the distinct dependence of the small-junction and array matrix elements on the external flux [Fig.~\ref{fig:overview}(b)]. Notably, the model yields $x_\mathrm{qp}^\mathrm{array} \ll x_\mathrm{qp}^\mathrm{small}$ for Q1--Q4 [Table~\ref{tab:xqp_models}], consistent with the disparity between the corresponding QP-density bounds reported in the literature [Table~\ref{tab:qp_density_summary}]. However, concluding that $x_\mathrm{qp}^\mathrm{array} \ll x_\mathrm{qp}^\mathrm{small}$ would be premature, as this is not supported after accounting for the relevant energy scales, namely $\hbar\omega_{01}$, $\delta E_\mathrm{qp}$, and $\delta\Delta$, as demonstrated in the main text.

\section{Quasiparticle-induced rates in the asymmetric superconducting gap model}
\label{app:asymmetric_gap_model}

\begin{figure*}
 \centering
 \includegraphics[width=\textwidth]{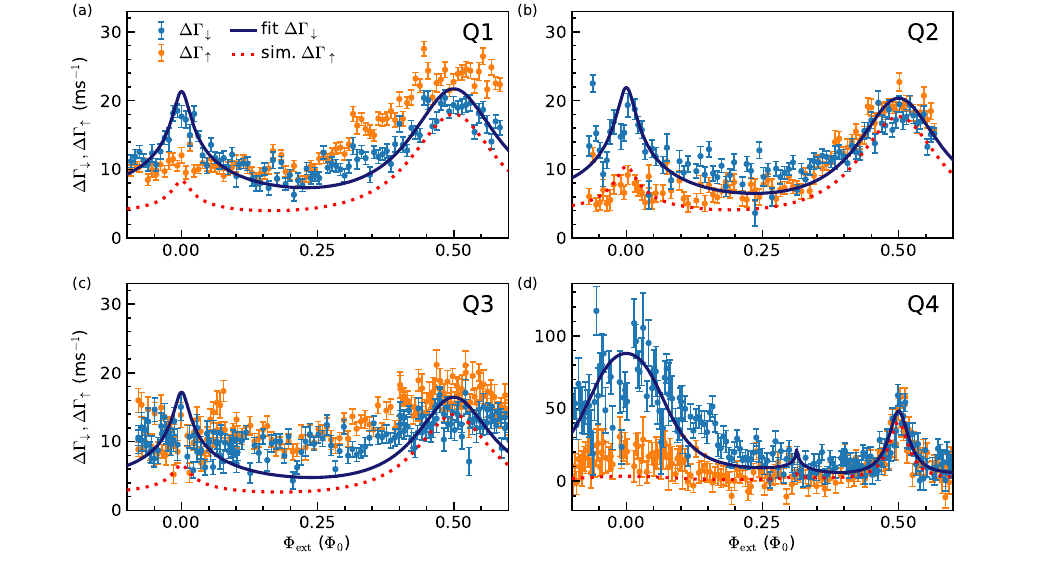}
 \caption{\textbf{Fits of the quasiparticle-induced de-excitation rates $\Delta\Gamma_\downarrow$ to a general model for qubits Q1--Q4.} (a)--(d)~QP-induced increase in the de-excitation ($\Delta\Gamma_\downarrow = \Gamma_\downarrow^\mathrm{inj} - \Gamma_\downarrow^\mathrm{bg}$) and excitation ($\Delta\Gamma_\uparrow = \Gamma_\uparrow^\mathrm{inj} - \Gamma_\uparrow^\mathrm{bg}$) rates as functions of the external flux $\Phi_\mathrm{ext}$ for qubits Q1, Q2, Q3, and Q4, respectively. The transition rates in qubits Q1--Q3 exhibit an apparent inversion ($\Delta\Gamma_\uparrow>\Delta\Gamma_\downarrow$) near the HFQ, while retaining their regular ordering ($\Delta\Gamma_\uparrow<\Delta\Gamma_\downarrow$) near the IFQ. The solid blue curves are the two-parameter least-squares fits to $\Delta\Gamma_\downarrow$, with the reduced QP density $x_\mathrm{qp}$ and effective QP temperature $T_\mathrm{qp}$ as fit parameters, while the red dotted curves show the corresponding model predictions for the QP-induced excitation rates $\Delta\Gamma_\uparrow$. The general model assumes $\delta\Delta/h = 1.72\,\mathrm{GHz}$ and $x_\mathrm{qp}^\mathrm{small} = x_\mathrm{qp}^\mathrm{array}$. Injection parameters are (a)~$V_\mathrm{inj} = 5.0\,\Delta_\mathrm{Al}/e$, $t_\mathrm{inj} = 4.0\,\mu\mathrm{s}$; (b)~$V_\mathrm{inj} = 12.0\,\Delta_\mathrm{Al}/e$, $t_\mathrm{inj} = 1.0\,\mu\mathrm{s}$; (c)~$V_\mathrm{inj} = 2.9\,\Delta_\mathrm{Al}/e$, $t_\mathrm{inj} = 1.0\,\mu\mathrm{s}$; (d)~$V_\mathrm{inj} = 6.1\,\Delta_\mathrm{Al}/e$, $t_\mathrm{inj} = 40.0\,\mu\mathrm{s}$. The effective QP temperatures obtained from the fits are (a)~$k_B T_\mathrm{qp}/h = 1.4\,\mathrm{GHz}$; (b)~$k_B T_\mathrm{qp}/h = 2.1\,\mathrm{GHz}$; (c)~$k_B T_\mathrm{qp}/h = 1.6\,\mathrm{GHz}$; (d)~$k_B T_\mathrm{qp}/h = 1.2\,\mathrm{GHz}$. The y-axis in panel (d) is adjusted to show the full range of measured transition rates for Q4.}
 \label{fig:xqp_fits}
\end{figure*}

QP-induced transition rates in qubits Q1--Q4 are measured using different injection parameters, so the corresponding QP energy distributions are likely to differ. For Q1, the effective QP temperature $T_\mathrm{qp}$ is independently estimated from the injection response of a transmon qubit co-fabricated on the same chip. From the measurements used to estimate $T_\mathrm{qp}$, we extract the reduced QP density $x_\mathrm{qp}^\mathrm{tr}$ while accounting for $\delta\Delta \neq 0$. To this end, we retain only the corresponding small junction terms in Eq.~\eqref{eq:rates}, with the QP tunneling structure factors of Eq.~\eqref{eq:s_qp} and the thermal QP distribution function of Eq.~\eqref{eq:f_th}. After estimating $T_\mathrm{qp}$ from the detailed balance relation, we numerically solve the equation relating the measured transmon de-excitation rate to the QP density, $\Delta\Gamma_\downarrow^\mathrm{tr} = \Delta\Gamma_\downarrow(x_\mathrm{qp}^\mathrm{tr})$. We use $\delta\Delta/h = 1.72\,\mathrm{GHz}$ and $k_B T_\mathrm{qp}/h = 1.2\,\mathrm{GHz}$ in this equation. For injection parameters $V_\mathrm{inj} = 5.0 \,\Delta_\mathrm{Al}/e$ and $t_\mathrm{inj} = 4.0 \,\mu\mathrm{s}$, we obtain $x_\mathrm{qp}^\mathrm{tr}=3.0 \times {10^{-6}}$.

Because no corresponding transmon data are available for Q2--Q4, we generalize the one-parameter fit in Sec.~\ref{sec:de-excitation_asymmetric_case} by treating both $x_\mathrm{qp}$ and $T_\mathrm{qp}$ as fit parameters. For consistency, we perform this two-parameter fit for all qubits Q1--Q4. Figure~\ref{fig:xqp_fits} shows two-parameter fits (blue solid lines) to the measured QP-induced de-excitation rates $\Delta\Gamma_\downarrow$ for qubits Q1--Q4 using the general $\delta\Delta \neq 0$ model. The model is presented in Sec.~\ref{sec:de-excitation_asymmetric_case} and defined by Eqs.~\eqref{eq:rates}, \eqref{eq:s_qp}, and~\eqref{eq:f_th}. We assume the same QP distribution in the small junction and in the junction array with a single reduced QP density $x_\mathrm{qp}$ and effective temperature $T_\mathrm{qp}$ in Eq.~\eqref{eq:f_th}. The extracted values of $x_\mathrm{qp}$ are reported in Table~\ref{tab:xqp_models}, and the corresponding values of $T_\mathrm{qp}$ are given in the caption of Fig.~\ref{fig:xqp_fits}. The fits capture the main features of $\Delta\Gamma_\downarrow$ while qualitatively accounting for the nonzero excitation rates $\Delta\Gamma_\uparrow$. The variation in the fitted $T_\mathrm{qp}$ values likely arises because the QP energy distribution following injection, and hence the corresponding effective QP temperature, depends on the injection parameters. We attribute the remaining discrepancies between the measured and simulated rates to leakage to higher excited states and to potential deviations of the QP energy distribution from the assumed Boltzmann form~\cite{Martinis2009, Marchegiani2025, KurilovichV2026}, as discussed in detail in Sec.~\ref{sec:excitation_rate} and~Appendix~\ref{app:rate_inversion}.

\begin{table}[t]
\centering
\small
\renewcommand{\arraystretch}{1.2}
\begin{tabular}{c@{\hspace{12pt}}ccc}
\toprule
& \multicolumn{2}{c}{%
    \shortstack{Simplified model\\
    of Appendix~\ref{app:simple_model}}}
& \shortstack{$\delta\Delta \neq 0$ model\\
    of Appendix~\ref{app:asymmetric_gap_model}} \\
\cmidrule(lr){2-3}
\cmidrule(lr){4-4}

Qubit
& $x_\mathrm{qp}^\mathrm{small}$ ($10^{-6}$)
& $x_\mathrm{qp}^\mathrm{array}$ ($10^{-6}$)
& $x_\mathrm{qp}$ ($10^{-6}$) \\
\midrule

Q1 & 2.1 & 0.4 & 3.1 \\
Q2 & 1.8 & 0.3 & 2.0 \\
Q3 & 2.0 & 0.3 & 2.0 \\
Q4 & 3.2 & 0.3 & 3.3 \\

\bottomrule
\end{tabular}
\caption{\textbf{Reduced quasiparticle densities $x_\mathrm{qp}$ extracted from fits to the de-excitation transition rates $\Delta\Gamma_\downarrow$.} The first two columns report $x_\mathrm{qp}^\mathrm{small}$ and $x_\mathrm{qp}^\mathrm{array}$ obtained using the simplified model of Appendix~\ref{app:simple_model} under the commonly used assumptions $\delta\Delta=0$ and $\delta E_\mathrm{qp}\ll\hbar\omega_{01}$ [Fig.~\ref{fig:simple_xqp_fits}]. This model yields \mbox{$x_\mathrm{qp}^\mathrm{array}\ll x_\mathrm{qp}^\mathrm{small}$} for all qubits Q1--Q4. The third column reports $x_\mathrm{qp}$ extracted using the $\delta\Delta \neq 0$ model of Sec.~\ref{sec:de-excitation_asymmetric_case}, but with $x_\mathrm{qp}$ and $T_\mathrm{qp}$ treated as fit parameters, as described in Appendix~\ref{app:asymmetric_gap_model}. This model, which assumes the same QP distribution in the small junction and in the junction array, quantitatively captures the main features of $\Delta\Gamma_\downarrow$ while also accounting for the nonzero excitation rates $\Delta\Gamma_\uparrow$ [Fig.~\ref{fig:xqp_fits}].}
\label{tab:xqp_models}
\end{table}

\section{Three-level transition rate model}
\label{app:three_level_rate_model}

In this section, we extend the conventional two-level transition rate analysis to a three-level model. In principle, the full $N$-level transition rate matrix can be determined by preparing $N$ distinct initial population distributions and measuring the resulting qubit population dynamics. Experimentally, however, this approach faces challenges because the state-dependent dispersive shifts can be small, limiting the separation between the corresponding qubit states in the $I$--$Q$ plane, and the reliable preparation of higher excited states requires additional calibration. In our experiments on Q1, the state $\ket{2}$ is well resolved. We take advantage of this separation to analyze the population dynamics in states $\ket{0}$, $\ket{1}$, and $\ket{2}$ using a three-level transition rate model.

\subsection{Three-level population dynamics}
\label{app:three_level_rates}

\begin{figure}
 \centering
 \includegraphics[width=\columnwidth]{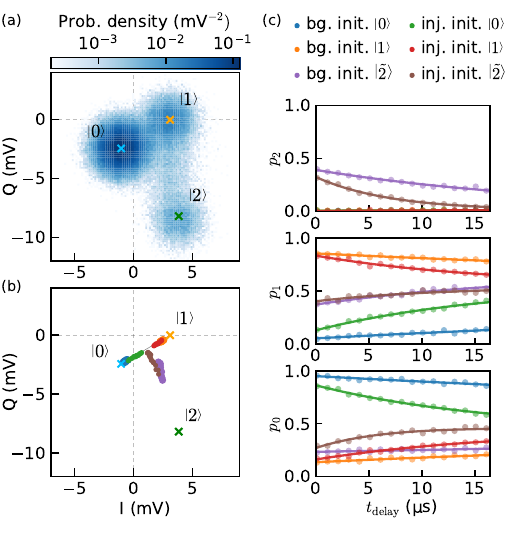}
 \caption{\textbf{Three-level population dynamics in qubit~Q1.} (a)~Measured single-shot readout distribution in the $I$--$Q$ plane at the readout flux bias $\Phi_\mathrm{ext}^\mathrm{readout} = 0.5\,\Phi_0$, following $0.8\,\mathrm{ms}$ of idling at $\Phi_\mathrm{ext} = 0.0\,\Phi_0$. The cyan, orange, and green crosses mark the centroids of the clusters corresponding to $\ket{0}$, $\ket{1}$, and $\ket{2}$, respectively. The gray dashed lines indicate $I=0$ and $Q=0$. (b)~Evolution of the averaged resonator response in the $I$--$Q$ plane following three distinct initial-state preparations for the background and injection measurements. The blue, orange, and purple points represent background measurements of the averaged resonator response following qubit preparation in three distinct qubit population distributions labeled as initialization in $\ket{0}$, $\ket{1}$, and $\ket{\tilde{2}}$, respectively. The $\ket{\tilde{2}}$ label denotes a distribution with substantial population in state $\ket{2}$, hence the label choice. The green, red, and brown points represent the corresponding trajectories in the QP-injection measurements. $V_\mathrm{inj} = 7.0\,\Delta_\mathrm{Al}/e$, $t_\mathrm{inj} = 2.0\,\mu\mathrm{s}$, and $N=2000$. (c)~Populations $p_0$, $p_1$, and $p_2$ as functions of $t_\mathrm{delay}$ for the data shown in panel~(b). The points are color-coded according to the initial-state preparation and the measurement type (background or injection). Solid lines show fits to the three-level transition rate model. The transition rates are extracted from local joint fits applied separately to the background (blue, orange, and purple points) and the injection (green, red, and brown points) data over the time window $0 \leq t_\mathrm{delay} \leq 16\,\mu\mathrm{s}$. Extracted background transition rates: $\Gamma_{10}^\mathrm{bg} = 6.4 \,\mathrm{ms^{-1}}$, $\Gamma_{01}^\mathrm{bg} = 6.1 \,\mathrm{ms^{-1}}$, $\Gamma_{20}^\mathrm{bg} = 1.3 \,\mathrm{ms^{-1}}$, $\Gamma_{02}^\mathrm{bg} = 0.0 \,\mathrm{ms^{-1}}$, $\Gamma_{21}^\mathrm{bg} = 42.0 \,\mathrm{ms^{-1}}$, and $\Gamma_{12}^\mathrm{bg} = 0.7 \,\mathrm{ms^{-1}}$. Extracted post-injection transition rates: $\Gamma_{10}^\mathrm{inj} = 25.5 \,\mathrm{ms^{-1}}$, $\Gamma_{01}^\mathrm{inj} = 33.5 \, \mathrm{ms^{-1}}$, $\Gamma_{20}^\mathrm{inj} = 97.4 \,\mathrm{ms^{-1}}$, $\Gamma_{02}^\mathrm{inj} = 1.1 \,\mathrm{ms^{-1}}$, $\Gamma_{21}^\mathrm{inj} = 43.8 \,\mathrm{ms^{-1}}$, and $\Gamma_{12}^\mathrm{inj} = 1.9 \,\mathrm{ms^{-1}}$.}
 \label{fig:q1_three_level_hfq}
\end{figure}

Figure~\ref{fig:q1_three_level_hfq}(a) shows the readout outcome clusters in the $I$--$Q$ plane corresponding to the qubit states $\ket{0}$, $\ket{1}$, and $\ket{2}$. The separation between clusters arises from the qubit-state-dependent dispersive shifts $\chi_i$ of the bare resonator frequency $\omega_\mathrm{res}$ [Fig.~\ref{fig:spectra}]. In the three-level model, the averaged response of the resonator $S_{21}$ is given by $S_{21}(t) = \sum_{i=0}^{2}p_i(t)S_{21}^{(i)}$, where $p_i(t)$ denotes the population of state $\ket{i}$ and $S_{21}^{(i)}$ is the centroid of the corresponding state cluster in the $I$--$Q$ plane (we omit a~calibration factor that relates the value of the $S_{21}$ parameter to the demodulated voltage $V_\mathrm{ADC} = I+iQ$ that defines a point in the $I$--$Q$ plane). The population vector $\mathbf{p}(t)=\bigl[p_0(t),p_1(t),p_2(t)\bigr]^{\mathsf{T}}$ satisfies the normalization condition $\sum_{i=0}^{2}p_i(t)=1$.

To determine the three-level transition matrix, we first measure the population dynamics at $\Phi_\mathrm{ext} = 0.5\,\Phi_0$ for three distinct initial qubit populations [Fig.~\ref{fig:q1_three_level_hfq}(b)]. In addition to initializing the qubit in $\ket{0}$ and $\ket{1}$, we use an initial population distribution that contains an appreciable population in $\ket{2}$, which we denote as $\ket{\tilde{2}}$. This initial distribution is prepared by first initializing the qubit in $\ket{1}$ and then rapidly tuning the external flux from $0.5\,\Phi_0$ to $0.0\,\Phi_0$ and back to $0.5\,\Phi_0$, thus implementing a diabatic flux drive pulse. At $\Phi_\mathrm{ext} = 0.0\,\Phi_0$, the transition frequency $\omega_{12}$ of Q1 is relatively small, allowing population transfer to $\ket{2}$. We then measure the averaged resonator response $S_{21}(t)$ and determine the time evolution of the population vector $\mathbf{p}(t)$ using the centroid positions $S_{21}^{(i)}$ [Fig.~\ref{fig:q1_three_level_hfq}(c)].

The time evolution of the population vector $\mathbf{p}(t)$ is governed by the rate equation $\dot{\mathbf{p}}(t) = \boldsymbol{\Gamma}\mathbf{p}(t)$, where the three-level transition rate matrix is given by
\begin{equation}
  \begingroup
  \setlength{\arraycolsep}{-2pt}
  \boldsymbol{\Gamma}  =
  \begin{pmatrix}
    -(\Gamma_{01}+\Gamma_{02}) & \Gamma_{10} & \Gamma_{20} \\
    \Gamma_{01} & -(\Gamma_{10}+\Gamma_{12}) & \Gamma_{21} \\
    \Gamma_{02} & \Gamma_{12} & -(\Gamma_{20}+\Gamma_{21})
  \end{pmatrix},
  \endgroup
\end{equation}
where $\Gamma_{if}$ denotes the transition rate from state $\ket{i}$ to state $\ket{f}$. For a given initial population vector $\mathbf{p}(0)$, the solution of the rate equation is
\begin{equation}
    \mathbf{p}(t) = e^{\boldsymbol{\Gamma}t}\mathbf{p}(0).
\end{equation}
Finally, we perform a local joint fit of the inferred populations to the solution of the rate equation, separately fitting the background and injection data, using rates $\Gamma_{if}$ and the three initial population vectors $\mathbf{p}^{k}(0)$, $k\in\{\ket{0},\ket{1},\ket{\tilde{2}}\}$, as fitting parameters. Representative results of this three-level analysis are shown in Fig.~\ref{fig:q1_three_level_hfq}(c).

\subsection{External flux dependence of the transition rates}
\label{app:three_level_rates_vs_flux}

\begin{figure*}
 \centering
 \includegraphics[width=\textwidth]{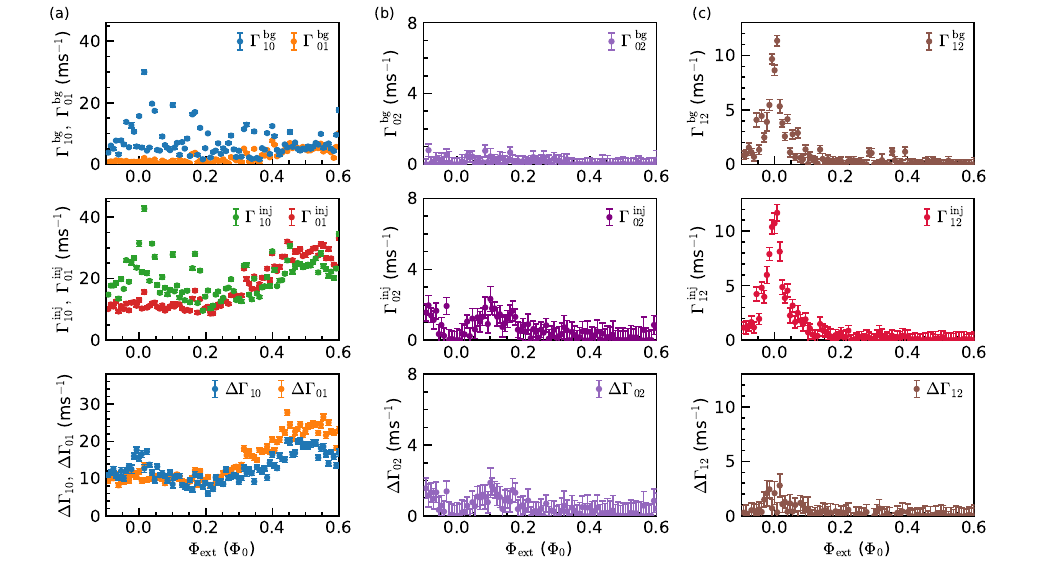}
 \caption{{\bf Transition rates extracted from the three-level model versus the external flux.} (a)~Transition rates $\Gamma_{10}$ and $\Gamma_{01}$ as functions of the external flux $\Phi_\mathrm{ext}$. (b)~Transition rate $\Gamma_{02}$ as a function of $\Phi_\mathrm{ext}$. (c)~Transition rate $\Gamma_{12}$ as a function of $\Phi_\mathrm{ext}$. In each column, the upper, middle, and lower panels show the background rates $\Gamma_{if}^\mathrm{bg}$, post-injection rates $\Gamma_{if}^\mathrm{inj}$, and QP-induced transition rates $\Delta\Gamma_{if} = \Gamma_{if}^\mathrm{inj}-\Gamma_{if}^\mathrm{bg}$, respectively. The transition rates are extracted from joint fits of the measured qubit populations over the time window $0 \leq t_\mathrm{delay} \leq 16\,\mu\mathrm{s}$ to the three-level model, with separate fits performed for the background and injection measurements. The injection parameters are $V_\mathrm{inj} = 5.0\,\Delta_\mathrm{Al}/e$ and $t_\mathrm{inj} = 4.0\,\mu\mathrm{s}$, which are the same as those in Fig.~\ref{fig:rates_ext_flux} of the main text.}
 \label{fig:q1_three_level_rates_vs_flux}
\end{figure*}

In this section, we revisit the analysis of the data presented in Fig.~\ref{fig:rates_ext_flux} using the three-level transition rate model. In this dataset, the qubit was initialized only in $\ket{0}$ and $\ket{1}$, which is not sufficient to fully constrain the three-level transition rate matrix. However, assuming that these initializations contain a negligible initial population in $\ket{2}$, the terms proportional to $\Gamma_{20}$ and $\Gamma_{21}$ do not contribute to the initial population dynamics to linear order in time. We therefore restrict the analysis to the four transition rates $\Gamma_{01}$, $\Gamma_{02}$, $\Gamma_{10}$, and $\Gamma_{12}$.

Figure~\ref{fig:q1_three_level_rates_vs_flux} shows the external flux dependence of the three-level qubit transition rates for the background, post-injection, and QP-induced contributions. We find that the three-level analysis does not substantially alter the extracted $\Delta\Gamma_{01}$ and $\Delta\Gamma_{10}$ transition rates compared to those presented in Fig.~\ref{fig:rates_ext_flux}. In particular, the apparent inversion of the QP-induced transition rates near the HFQ is preserved, while the rates retain their regular ordering near the IFQ. As a consistency check, we refit the model of Sec.~\ref{sec:de-excitation_asymmetric_case} to the three-level de-excitation data $\Delta\Gamma_{10}$ using the experimentally inferred $\delta\Delta$ and $T_\mathrm{qp}$, with a common QP density $x_\mathrm{qp}$ as the fit parameter. We obtain $x_\mathrm{qp} = 3.6\times10^{-6}$, consistent with the value obtained from fitting the two-level de-excitation rates. This agreement indicates that including the third level in the rate-extraction analysis has only a minor effect on the transition rates within the computational subspace $\{\ket{0},\ket{1}\}$ for Q1.

Next, we discuss the QP-induced transition rates $\Delta\Gamma_{02}$ and $\Delta\Gamma_{12}$ [Figs.~\ref{fig:q1_three_level_rates_vs_flux}(b,\,c)]. Remarkably, $\Delta\Gamma_{02}$ exhibits a peak near $\Phi_\mathrm{ext} \approx 0.104\,\Phi_0$, where the transition frequency is $\omega_{02}/2\pi \approx 1.72\,\mathrm{GHz}$. Within the QP-induced qubit transition rate model, such an enhancement is expected when the qubit transition energy matches the difference between the SC gaps of the two Josephson junction leads, i.e., when $\hbar\omega_{if}=\delta\Delta$~\cite{Diamond2022, Marchegiani2022}. This resonant response in the transition rate arises from enhanced QP tunneling due to an increased QP density of states near the SC gap edges, thereby facilitating faster energy exchange between the qubit and QPs. Using the BCS density of states, the theoretical peak in Fig.~\ref{fig:q1_rates_to_higher_levels}(a) is logarithmically divergent; in practice, however, this divergence is cut off by mechanisms such as the proximity effect~\cite{Marchegiani2025arXiv}. Therefore, the position of the peak in $\Delta\Gamma_{02}$ provides a direct estimate of the difference in the SC gaps: $\delta\Delta/h \approx 1.72\,\mathrm{GHz}$. This value is in good agreement with the estimate $\delta\Delta/h \approx 1.8\,\mathrm{GHz}$ obtained from the dependence of the SC gap on the thickness of the Al film [Appendix~\ref{app:qubits}]. Finally, we observe an enhancement of $\Delta\Gamma_{12}$ near the IFQ point, which we attribute to the small transition frequency $\omega_{12}$ in this flux region. In fact, at the IFQ, $\omega_{12}/2\pi \simeq 90\,\mathrm{MHz} \ll k_B T_\mathrm{qp}/h \sim \delta\Delta/h$.

\section{Apparent rate inversion}
\label{app:rate_inversion}

In this section, we discuss two possible scenarios for the apparent inversion of QP-induced transition rates $\Delta\Gamma_\uparrow > \Delta\Gamma_\downarrow$ observed for qubit Q1 [Fig.~\ref{fig:rates_ext_flux}(c)]. A similar effect is observed in qubits Q2 and Q3 [Fig.~\ref{fig:xqp_fits}]. Although the general discussion applies to these qubits as well, the analysis in this section focuses on Q1.

\subsection{Leakage to higher excited states}
\label{app:leakage_to_noncomp_states}

\begin{figure}
 \includegraphics[width=\columnwidth]{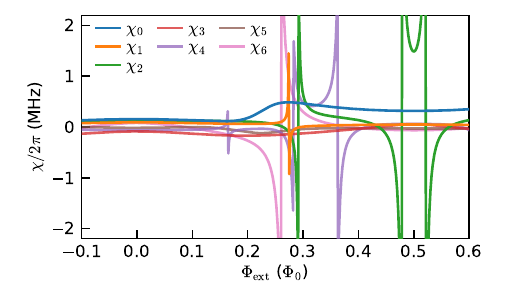}
 \caption{\textbf{State-dependent dispersive shifts in qubit Q1.}
  Dispersive shifts $\chi_i$ of the resonator frequency for the first seven fluxonium states $\{\ket{0},\ket{1},\dots,\ket{6}\}$ of qubit Q1 as functions of the external flux $\Phi_\mathrm{ext}$ obtained from the exact numerical diagonalization of the Hamiltonian~\eqref{eq:H_system} with the parameters listed in Table~\ref{tab:qubit_parameters}. At the HFQ, the states $\ket{0}$, $\ket{1}$, and $\ket{2}$ exhibit distinct dispersive shifts, resulting in well-separated readout clusters in the $I$--$Q$ plane [Fig.~\ref{fig:q1_three_level_hfq}(a)]. In contrast, the excited states $\ket{3}$, $\ket{4}$, $\ket{5}$, and $\ket{6}$ exhibit dispersive shifts close to $\chi_1$, causing the readout clusters associated with these states and state $\ket{1}$ to overlap.}
 \label{fig:q1_chi_shifts}
\end{figure}

At the readout flux bias $\Phi_\mathrm{ext}^\mathrm{readout} = 0.5\,\Phi_0$, Figure~\ref{fig:q1_three_level_hfq}(a) shows that the Q1 readout clearly resolves three clusters corresponding to $\ket{0}$, $\ket{1}$, and $\ket{2}$. Their separation arises from the distinct dispersive shifts $\chi_0$, $\chi_1$, and $\chi_2$ of the bare resonator frequency, in good agreement with our simulations [Fig.~\ref{fig:q1_chi_shifts}]. We also simulate the dispersive shifts for a few higher states. The simulations predict that the dispersive shifts of $\{\ket{3},\ket{4},\ket{5},\ket{6}\}$ lie close to that of $\ket{1}$, causing their readout clusters to overlap with the $\ket{1}$ cluster. Therefore, those higher excited states would be misidentified as $\ket{1}$.

There are three stages in our measurement protocol shown in Fig.~\ref{fig:protocol}(a) during which the qubit population can transfer to higher excited states, resulting in state misidentification: (i)~QP-induced population transfer during the injection pulse of duration $t_\mathrm{inj}$ at the readout flux bias~$\Phi_\mathrm{ext}^\mathrm{readout}$, (ii)~QP-induced dynamics at the operating flux bias $\Phi_\mathrm{ext}$ during the delay time $t_\mathrm{delay}$, and (iii)~QP-induced dynamics during the readout time $t_\mathrm{readout}$ at $\Phi_\mathrm{ext}^\mathrm{readout}$.

\begin{figure}
 \includegraphics[width=\columnwidth]{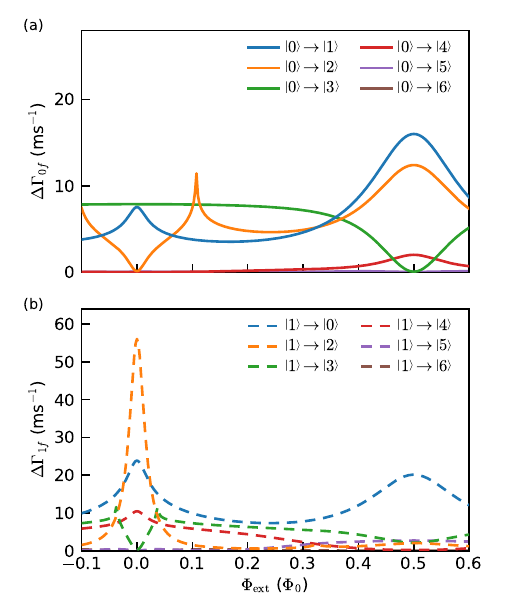}
 \caption{\textbf{Simulated QP-induced transition rates from states $\ket{0}$ and $\ket{1}$ for qubit Q1.} (a)~QP-induced transition rates $\Delta\Gamma_{0f}$ for transitions from $\ket{0}$ to $\ket{f}$, $f \in \{1,\ldots,6\}$. (b)~QP-induced transition rates $\Delta\Gamma_{1f}$ for transitions from $\ket{1}$ to $\ket{f}$, $f \in \{0,2,\ldots,6\}$. Transitions to higher excited states can cause leakage from the $\{\ket{0},\ket{1}\}$ subspace and, when these higher excited states are misidentified during readout, modify the extracted rates relative to the true qubit transition rates. The model parameters are the same as in Fig.~\ref{fig:rates_ext_flux} of the main text.}
 \label{fig:q1_rates_to_higher_levels}
\end{figure}

To extract $\Delta\Gamma_\uparrow$ and $\Delta\Gamma_\downarrow$, we initialize the qubit predominantly in $\ket{0}$ and $\ket{1}$, respectively. The initial post-injection population dynamics is therefore governed primarily by transitions out of the initialized states. Our model of QP-induced transition rates in Sec.~\ref{sec:de-excitation_asymmetric_case}, which accounts for a finite SC gap difference $\delta\Delta$ between the junction leads and a finite QP temperature $T_\mathrm{qp}$, predicts that the transition rates from $\ket{0}$ to $\ket{2}$, $\ket{3}$, and $\ket{4}$ are comparable to the $\ket{0}\!\to\!\ket{1}$ transition rate [Fig.~\ref{fig:q1_rates_to_higher_levels}]. Because the population in $\ket{3}$, $\ket{4}$, and other higher excited states is misidentified as the population in $\ket{1}$, these transitions increase the apparent excitation rate relative to the true $\ket{0}\!\to\!\ket{1}$ rate. By contrast, the transitions from $\ket{1}$ to $\ket{3}$--$\ket{6}$ leave the population within the readout cluster associated with $\ket{1}$. Consequently, the apparent population transfer from the $\ket{1}$ cluster to $\ket{0}$ largely reflects the true de-excitation rate. This motivates our emphasis in the main text on the theoretical analysis of $\Delta\Gamma_\downarrow$ rather than $\Delta\Gamma_\uparrow$, which is more strongly affected by leakage to higher excited states.

We note that explicitly including the $\ket{2}$ state in the rate extraction analysis does not significantly change the observed apparent transition rates $\Delta\Gamma_\downarrow$ and $\Delta\Gamma_\uparrow$ [Appendix~\ref{app:three_level_rates_vs_flux}]. Nevertheless, in the presence of rapid QP-induced population dynamics at the readout flux bias $\Phi_\mathrm{ext}^\mathrm{readout} = 0.5\,\Phi_0$, we cannot exclude partial misidentification of $\ket{2}$ due to transitions occurring during readout, such as $\ket{2}\!\to\!\ket{0}$ and $\ket{2}\!\to\!\ket{1}$, as well as other transitions out of $\ket{2}$. Such processes could cause population in $\ket{2}$ to be assigned to the $\ket{0}$ or $\ket{1}$ readout clusters, thereby biasing the apparent transition rates even within the three-level model. In fact, our simulations predict larger values of $\Delta\Gamma_{02}$ and $\Delta\Gamma_{12}$ than those observed experimentally [Fig.~\ref{fig:q1_rates_to_higher_levels}], suggesting that a fraction of the $\ket{2}$ population is misidentified as $\ket{0}$ or $\ket{1}$.

Finally, we do not exclude the possibility of measurement-induced state transitions (MIST)~\cite{Bista2026, Singh2025, Chapple2026, Zwanenburg2026}, which are not included in our analysis and could modify the apparent population dynamics. However, we do not expect this effect to be dominant because a substantial MIST contribution would likely produce a similar apparent rate inversion even in the absence of QP injection, whereas no such inversion is observed in the background measurements.

We therefore identify leakage to higher excited states as a plausible explanation for the observed apparent rate inversion. Distinguishing and quantifying the effects of leakage would require improved state preparation during QP injection, shorter readout times, and high-fidelity readout of higher qubit states.

\subsection{Alternative quasiparticle energy distribution}
\label{app:alternative_qp_dist}

\begin{figure}
 \includegraphics[width=\columnwidth]{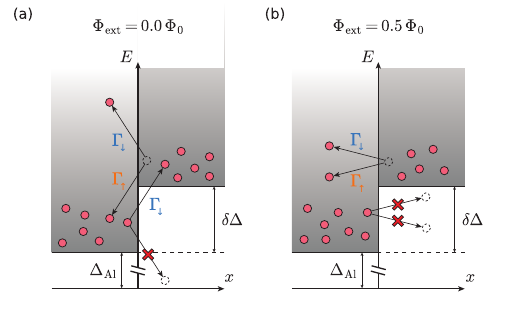}
 \caption{\textbf{Schematic of QP-induced qubit excitation and de-excitation processes in an asymmetric Josephson junction.} (a)~Near the IFQ, the larger qubit transition energy allows QPs in the low-gap lead to absorb energy from the qubit and tunnel into the high-gap lead, favoring the regular QP-induced rate ordering $\Gamma_\uparrow < \Gamma_\downarrow$. (b)~Near the HFQ, the dominant sine-term contribution of the small junction is suppressed, and the transition rates are dominated by QP tunneling in the junction array. For a non-equilibrium QP distribution with characteristic width $\delta E_\mathrm{qp} \ll \delta\Delta$, the QPs in the high-gap lead preferentially excite the qubit due to the higher final density of states for the QPs in these processes, whereas the near-gap QPs in the low-gap lead do not contribute appreciably to the transition rates, resulting in the QP-induced rate inversion $\Gamma_\uparrow > \Gamma_\downarrow$. The SC gap difference $\delta\Delta$ offsets the QP density of states in the low- and high-gap leads. Darker gray shading corresponds to a higher density of states. Red circles depict non-equilibrium QPs, and arrows denote inelastic QP tunneling processes accompanied by energy exchange with the qubit.}
 \label{fig:qp_transition_schematic}
\end{figure}

\begin{figure}
 \includegraphics[width=\columnwidth]{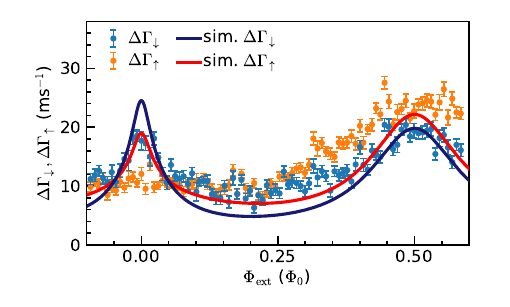}
 \caption{\textbf{Comparison of QP-induced transition rates measured for Q1 to a simulation with a narrow athermal QP energy distribution.} QP-induced de-excitation $\Delta\Gamma_\downarrow$ (blue points) and excitation $\Delta\Gamma_\uparrow$ (orange points) transition rates as functions of external flux~$\Phi_\mathrm{ext}$ [same as in Fig.~\ref{fig:rates_ext_flux}(c)]. The solid blue and red curves show numerical simulations of the QP-induced de-excitation and excitation rates, respectively, using a model with a narrow athermal QP distribution described in Appendix~\ref{app:alternative_qp_dist}. The model predicts an inversion of the QP-induced transition rates $\Delta\Gamma_\uparrow > \Delta\Gamma_\downarrow$ near the HFQ while retaining the regular ordering $\Delta\Gamma_\uparrow < \Delta\Gamma_\downarrow$ near the IFQ. The model parameters are $\delta\Delta/h \approx 1.72\,\mathrm{GHz}$, $k_B T_\mathrm{qp}/h = 0.20\,\mathrm{GHz}$, and $\delta\mu/\delta\Delta = 0.81$.}
 \label{fig:alternative_model}
\end{figure}

To entertain the alternative possibility that the apparent rate inversion arises from a strongly athermal QP energy distribution, characterized by different chemical potentials in the low- and high-gap leads and a narrow energy width $\delta E_\mathrm{qp}$, we present simulations demonstrating that a genuine rate inversion is theoretically possible.

Figure~\ref{fig:qp_transition_schematic} illustrates the key processes in this model. At the HFQ, where the qubit frequency is low, this model predicts a rate inversion because QPs that excite the qubit tunnel from the high-gap lead into a larger number of available final states, whereas QPs that de-excite the qubit tunnel to fewer final states. This asymmetry arises from the energy dependence of the BCS density of states. In contrast, the near-gap QPs in the low-gap lead cannot excite or de-excite the qubit because no final states are available below the gap of the opposite lead. This mechanism requires a narrow characteristic width of the QP distribution, $\delta E_\mathrm{qp} \ll \delta\Delta - \hbar\omega_{01}$; otherwise, QPs in the low-gap lead would contribute substantially to qubit de-excitation and suppress the rate inversion at the HFQ.

As a proxy for the QP distribution function, we use a~Boltzmann form with effective QP temperature $T_\mathrm{qp}$. To phenomenologically account for an athermal QP distribution, we introduce a chemical potential offset $\delta\mu$ between the low- and high-gap leads, analogous to the introduction of effective chemical potentials as in the ``local quasi-equilibrium'' regime of Ref.~\cite{Marchegiani2025}. Distinguishing between QPs in the small junction and in the array, as well as between QPs in the low- and high-gap leads, we write
\begin{equation}
f_{l}^{\alpha}(E) = 
x_\mathrm{qp,L}^\alpha \sqrt{\frac{\Delta_L}{2\pi k_B T_\mathrm{qp}}}
\exp\left(-\frac{E-\Delta_L-\mu_l}{k_B T_\mathrm{qp}}\right).
\label{eq:f_th_diff_chem_pot}
\end{equation}
Here, $l \in \{L,H\}$ labels the lead, and $\alpha \in \{\mathrm{small},\mathrm{array}\}$ labels the circuit element. $x_\mathrm{qp,L}^\alpha$ denotes the QP density on the low-gap side of the corresponding part of the circuit. $\mu_L=0$ and $\mu_H=\delta\mu$. For the simulation, we use the following parameters: $\delta\Delta/h = 1.72\,\mathrm{GHz}$, $k_B T_\mathrm{qp}/h = 0.20\,\mathrm{GHz}$, and $\delta\mu/\delta\Delta = 0.81$.

The simulated QP-induced transition rates under the assumption $x^\mathrm{small}_{\mathrm{qp}, \,l}=x^\mathrm{array}_{\mathrm{qp}, \,l}$ are in good qualitative agreement with the experimental data, as seen in Fig.~\ref{fig:alternative_model}: they capture the inversion of the rates at the HFQ, the crossover between the excitation and de-excitation rates, and the regular rate ordering at the IFQ. The resulting QP densities are $x^\alpha_{\mathrm{qp},L} \approx 1.0 \times 10^{-5}$ and
$x^\alpha_{\mathrm{qp},H} \approx 1.9 \times 10^{-6}$, $\alpha \in \{\mathrm{small},\mathrm{array}\}$, which gives~$x^\alpha_{\mathrm{qp},L}/x^\alpha_{\mathrm{qp},H} \approx 5.2$. These values are consistent with a picture in which QPs relax and subsequently accumulate in the low-gap lead~\cite{Marchegiani2022}, leading to $x_{\mathrm{qp},L} > x_{\mathrm{qp},H}$. For comparison, in a transmon with a larger gap difference than that in our samples, $\delta\Delta/h \simeq 5.5\,\mathrm{GHz}$, the QP density in the low-gap lead has been estimated to be about one order of magnitude larger than that in the high-gap lead~\cite{Krause2024}.

However, this model requires both a rather narrow QP energy distribution above the gap edge, with $\delta E_\mathrm{qp}/h \simeq 0.2\,\mathrm{GHz}$, and a highly athermal QP distribution in which the chemical potentials in the low- and high-gap leads are offset by $\delta\mu \simeq 0.8\,\delta\Delta$. This interpretation is inconsistent with the broader QP energy distribution $\delta E_\mathrm{qp}/h \approx 1.2\,\mathrm{GHz}$ inferred independently from the injection response of the co-fabricated transmon qubit under the same injection parameters [Sec.~\ref{sec:qp_energy_distribution}]. Moreover, relaxation of the injected QPs to such low energies is expected to occur on timescales exceeding milliseconds~\cite{Glazman2021}, far longer than the timescale corresponding to our measurements.

\bibliography{literature}

@PREAMBLE{
 "\providecommand{\noopsort}[1]{}" 
 # "\providecommand{\singleletter}[1]{#1}%" 
}

@article{Acharya2025,
  title   = {Quantum error correction below the surface code threshold},
  author  = {Acharya, Rajeev and Abanin, Dmitry A. and Aghababaie-Beni, Laleh and Aleiner, Igor and Andersen, Trond I. and Ansmann, Markus and Arute, Frank and Arya, Kunal and Asfaw, Abraham and Astrakhantsev, Nikita and others},
  journal = {Nature},
  year    = {2025},
  volume  = {638},
  issue   = {8052},
  pages   = {920--926},
  doi     = {10.1038/s41586-024-08449-y},
  url     = {https://doi.org/10.1038/s41586-024-08449-y}
}

@article{Wilen2021,
  title   = {Correlated charge noise and relaxation errors in superconducting qubits},
  author  = {Wilen, C. D. and Abdullah, S. and Kurinsky, N. A. and Stanford, C. and Cardani, L. and D’Imperio, G. and Tomei, C. and Faoro, L. and Ioffe, L. B. and Liu, C. H. and others},
  journal = {Nature},
  year    = {2021},
  volume  = {594},
  issue   = {7863},
  pages   = {369--373},
  doi     = {10.1038/s41586-021-03557-5},
  url     = {https://doi.org/10.1038/s41586-021-03557-5}
}

@article{McEwen2022,
  title   = {Resolving catastrophic error bursts from cosmic rays in large arrays of superconducting qubits},
  author  = {McEwen, Matt and Faoro, Lara and Arya, Kunal and Dunsworth, Andrew and Huang, Trent and Kim, Seon and Burkett, Brian and Fowler, Austin and Arute, Frank and Bardin, Joseph C. and others},
  journal = {Nature Physics},
  year    = {2022},
  volume  = {18},
  issue   = {1},
  pages   = {107--111},
  doi     = {10.1038/s41567-021-01432-8},
  url     = {https://doi.org/10.1038/s41567-021-01432-8}
}

@article{McEwen2024,
  title   = {Resisting High-Energy Impact Events through Gap Engineering in Superconducting Qubit Arrays},
  author  = {McEwen, Matt and Miao, Kevin C. and Atalaya, Juan and Bilmes, Alexander and Crook, Alex and Bovaird, Jenna and Kreikebaum, John Mark and Zobrist, Nicholas and Jeffrey, Evan and Ying, Bicheng and others},
  journal = {Phys. Rev. Lett.},
  year    = {2024},
  volume  = {133},
  issue   = {24},
  pages   = {240601},
  doi     = {10.1103/PhysRevLett.133.240601},
  url     = {https://link.aps.org/doi/10.1103/PhysRevLett.133.240601}
}

@article{Harrington2025,
  title = {Synchronous detection of cosmic rays and correlated errors in superconducting qubit arrays},
  author = {Harrington, Patrick M. and Li, Mingyu and Hays, Max and Van De Pontseele, Wouter and Mayer, Daniel and Pinckney, H. Douglas and Contipelli, Felipe and Gingras, Michael and Niedzielski, Bethany M. and Stickler, Hannah and others},
  journal = {Nature Communications},
  year = {2025},
  volume = {16},
  issue = {1},
  pages = {6428},
  doi = {10.1038/s41467-025-61385-x},
  url = {https://doi.org/10.1038/s41467-025-61385-x}
}

@article{KurilovichV2026,
  title   = {Correlated Phase Error Bursts in a Gap-Engineered Superconducting Qubit Array},
  author  = {Kurilovich, Vladislav D. and Roberts, Gabrielle and Martin, Leigh S. and McEwen, Matt and Eickbusch, Alec and Faoro, Lara and Ioffe, Lev B. and Atalaya, Juan and Bilmes, Alexander and Kreikebaum, John Mark and others},
  journal = {Phys. Rev. X},
  year    = {2026},
  volume  = {16},
  issue   = {2},
  pages   = {021025},
  doi     = {10.1103/1bl4-b2f7},
  url     = {https://link.aps.org/doi/10.1103/1bl4-b2f7}
}

@article{AnthonyPetersen2024,
  title   = {A stress-induced source of phonon bursts and quasiparticle poisoning},
  author  = {Anthony-Petersen, Robin and Biekert, Andreas and Bunker, Raymond and Chang, Clarence L. and Chang, Yen-Yung and Chaplinsky, Luke and Fascione, Eleanor and Fink, Caleb W. and Garcia-Sciveres, Maurice and Germond, Richard and others},
  journal = {Nature Communications},
  year    = {2024},
  volume  = {15},
  issue   = {1},
  pages   = {6444},
  doi     = {10.1038/s41467-024-50173-8},
  url     = {https://doi.org/10.1038/s41467-024-50173-8}
}

@article{Yelton2025,
  title   = {Correlated Quasiparticle Poisoning from Phonon-Only Events in Superconducting Qubits},
  author  = {Yelton, E. and Larson, C. P. and Dodge, K. and Okubo, K. and Plourde, B. L. T.},
  journal = {Phys. Rev. Lett.},
  year    = {2025},
  volume  = {135},
  issue   = {12},
  pages   = {123601},
  doi     = {10.1103/h65v-ttbw},
  url     = {https://link.aps.org/doi/10.1103/h65v-ttbw}
}

@article{Martinis2021,
  title   = {Saving superconducting quantum processors from decay and correlated errors generated by gamma and cosmic rays},
  author  = {Martinis, John M.},
  journal = {npj Quantum Information},
  year    = {2021},
  volume  = {7},
  issue   = {1},
  pages   = {90},
  doi     = {10.1038/s41534-021-00431-0},
  url     = {https://doi.org/10.1038/s41534-021-00431-0}
}

@article{Lutchyn2005,
  title   = {Quasiparticle decay rate of {Josephson} charge qubit oscillations},
  author  = {Lutchyn, Roman and Glazman, Leonid and Larkin, Anatoly},
  journal = {Phys. Rev. B},
  year    = {2005},
  volume  = {72},
  issue   = {1},
  pages   = {014517},
  doi     = {10.1103/PhysRevB.72.014517},
  url     = {https://link.aps.org/doi/10.1103/PhysRevB.72.014517}
}

@article{Martinis2009,
  title = {Energy Decay in Superconducting {Josephson}-Junction Qubits from Nonequilibrium Quasiparticle Excitations},
  author = {Martinis, John M. and Ansmann, M. and Aumentado, J.},
  journal = {Phys. Rev. Lett.},
  year = {2009},
  volume = {103},
  issue  = {9},
  pages = {097002},
  doi = {10.1103/PhysRevLett.103.097002},
  url = {https://link.aps.org/doi/10.1103/PhysRevLett.103.097002}
}

@article{Catelani2011,
  title   = {Relaxation and frequency shifts induced by quasiparticles in superconducting qubits},
  author  = {Catelani, G. and Schoelkopf, R. J. and Devoret, M. H. and Glazman, L. I.},
  journal = {Phys. Rev. B},
  year    = {2011},
  volume  = {84},
  issue   = {6},
  pages   = {064517},
  doi     = {10.1103/PhysRevB.84.064517},
  url     = {https://link.aps.org/doi/10.1103/PhysRevB.84.064517}
}

@article{Catelani2012,
  title   = {Decoherence of superconducting qubits caused by quasiparticle tunneling},
  author  = {Catelani, G. and Nigg, S. E. and Girvin, S. M. and Schoelkopf, R. J. and Glazman, L. I.},
  journal = {Phys. Rev. B},
  year    = {2012},
  volume  = {86},
  issue   = {18},
  pages   = {184514},
  doi     = {10.1103/PhysRevB.86.184514},
  url     = {https://link.aps.org/doi/10.1103/PhysRevB.86.184514}
}

@article{Iaia2022,
  title   = {Phonon downconversion to suppress correlated errors in superconducting qubits},
  author  = {Iaia, V. and Ku, J. and Ballard, A. and Larson, C. P. and Yelton, E. and Liu, C. H. and Patel, S. and McDermott, R. and Plourde, B. L. T.},
  journal = {Nature Communications},
  year    = {2022},
  volume  = {13},
  issue   = {1},
  pages   = {6425},
  doi     = {10.1038/s41467-022-33997-0},
  url     = {https://doi.org/10.1038/s41467-022-33997-0}
}

@article{Marchegiani2022,
  title   = {Quasiparticles in Superconducting Qubits with Asymmetric Junctions},
  author  = {Marchegiani, Giampiero and Amico, Luigi and Catelani, Gianluigi},
  journal = {PRX Quantum},
  year    = {2022},
  volume  = {3},
  issue   = {4},
  pages   = {040338},
  doi     = {10.1103/PRXQuantum.3.040338},
  url     = {https://link.aps.org/doi/10.1103/PRXQuantum.3.040338}
}

@misc{Pinckney2026,
  title         = {Characterization of Radiation-Induced Errors in Superconducting Qubits Protected with Various Gap-Engineering Strategies},
  author        = {Pinckney, H. Douglas and McJunkin, Thomas and Hunt, Alan W. and Harrington, Patrick M. and Binney, Hannah P. and Hays, Max and Jones-Alberty, Yenuel and Azar, Kate and Contipelli, Felipe and DePencier Piñero, Renée and others},
  year          = {2026},
  eprint        = {2603.13460},
  archivePrefix = {arXiv}
}

@article{Manucharyan2009,
  title   = {Fluxonium: Single {Cooper}-Pair Circuit Free of Charge Offsets},
  author  = {Manucharyan, Vladimir E. and Koch, Jens and Glazman, Leonid I. and Devoret, Michel H.},
  journal = {Science},
  year    = {2009},
  volume  = {326},
  issue   = {5949},
  pages   = {113--116},
  doi     = {10.1126/science.1175552},
  url     = {https://doi.org/10.1126/science.1175552}
}

@article{Nguyen2019,
  title = {High-Coherence Fluxonium Qubit},
  author = {Nguyen, Long B. and Lin, Yen-Hsiang and Somoroff, Aaron and Mencia, Raymond and Grabon, Nicholas and Manucharyan, Vladimir E.},
  journal = {Phys. Rev. X},
  year = {2019},
  volume = {9},
  issue = {4},
  pages = {041041},
  doi = {10.1103/PhysRevX.9.041041},
  url = {https://link.aps.org/doi/10.1103/PhysRevX.9.041041}
}

@article{Somoroff2023,
  title = {Millisecond Coherence in a Superconducting Qubit},
  author = {Somoroff, Aaron and Ficheux, Quentin and Mencia, Raymond A. and Xiong, Haonan and Kuzmin, Roman and Manucharyan, Vladimir E.},
  journal = {Phys. Rev. Lett.},
  year = {2023},
  volume = {130},
  issue = {26},
  pages = {267001},
  doi = {10.1103/PhysRevLett.130.267001},
  url = {https://link.aps.org/doi/10.1103/PhysRevLett.130.267001}
}

@article{Wang2025,
  title   = {High-coherence fluxonium qubits manufactured with a wafer-scale-uniformity process},
  author  = {Wang, Fei and Lu, Kannan and Zhan, Huijuan and Ma, Lu and Wu, Feng and Sun, Hantao and Deng, Hao and Bai, Yang and Bao, Feng and Chang, Xu and others},
  journal = {Phys. Rev. Appl.},
  year    = {2025},
  volume  = {23},
  issue   = {4},
  pages   = {044064},
  doi     = {10.1103/PhysRevApplied.23.044064},
  url     = {https://link.aps.org/doi/10.1103/PhysRevApplied.23.044064}
}

@article{Vool2014,
  title   = {Non-{Poissonian} Quantum Jumps of a Fluxonium Qubit due to Quasiparticle Excitations},
  author  = {Vool, U. and Pop, I. M. and Sliwa, K. and Abdo, B. and Wang, C. and Brecht, T. and Gao, Y. Y. and Shankar, S. and Hatridge, M. and Catelani, G. and others},
  journal = {Phys. Rev. Lett.},
  year    = {2014},
  volume  = {113},
  issue   = {24},
  pages   = {247001},
  doi     = {10.1103/PhysRevLett.113.247001},
  url     = {https://link.aps.org/doi/10.1103/PhysRevLett.113.247001}
}

@article{Pop2014,
  title   = {Coherent suppression of electromagnetic dissipation due to superconducting quasiparticles},
  author  = {Pop, Ioan M. and Geerlings, Kurtis and Catelani, Gianluigi and Schoelkopf, Robert J. and Glazman, Leonid I. and Devoret, Michel H.},
  journal = {Nature},
  year    = {2014},
  volume  = {508},
  issue   = {7496},
  pages   = {369--372},
  doi     = {10.1038/nature13017},
  url     = {https://doi.org/10.1038/nature13017}
}

@article{Grunhaupt2019,
  title   = {Granular aluminium as a superconducting material for high-impedance quantum circuits},
  author  = {Grünhaupt, Lukas and Spiecker, Martin and Gusenkova, Daria and Maleeva, Nataliya and Skacel, Sebastian T. and Takmakov, Ivan and Valenti, Francesco and Winkel, Patrick and Rotzinger, Hannes and Wernsdorfer, Wolfgang and others},
  journal = {Nature Materials},
  year    = {2019},
  volume  = {18},
  issue   = {8},
  pages   = {816--819},
  doi     = {10.1038/s41563-019-0350-3},
  url     = {https://doi.org/10.1038/s41563-019-0350-3}
}

@article{Atanasova2025,
  title   = {In Situ Tunable Interaction with an Invertible Sign between a Fluxonium and a Post Cavity},
  author  = {Atanasova, D. G. and Yang, I. and H\"onigl-Decrinis, T. and Gusenkova, D. and Pop, I. and Kirchmair, G.},
  journal = {PRX Quantum},
  year    = {2025},
  volume  = {6},
  issue   = {2},
  pages   = {020318},
  doi     = {10.1103/PRXQuantum.6.020318},
  url     = {https://link.aps.org/doi/10.1103/PRXQuantum.6.020318}
}

@misc{Ateshian2025,
  title         = {Temperature and Magnetic-Field Dependence of Energy Relaxation in a Fluxonium Qubit},
  author        = {Ateshian, Lamia and Hays, Max and Rower, David A. and Zhang, Helin and Azar, Kate and Assouly, R\'eouven and Ding, Leon and Gingras, Michael and Stickler, Hannah and Niedzielski, Bethany M. and others},
  year          = {2025},
  eprint        = {2507.01175},
  archivePrefix = {arXiv}
}

@misc{Azar2026Exp,
  title         = {Characterization and Comparison of Energy Relaxation in Fluxonium Qubits},
  author        = {Azar, Kate and Ateshian, Lamia and Randeria, Mallika T. and DePencier Piñero, Renée and Gertler, Jeffrey M. and An, Junyoung and Contipelli, Felipe and Ding, Leon and Gingras, Michael and Grossklaus, Kevin and others},
  year          = {2026},
  eprint        = {2603.23636},
  archivePrefix = {arXiv}
}

@article{Zhuang2026,
  title   = {{Non-Markovian} relaxation spectroscopy of fluxonium qubits},
  author  = {Zhuang, Ze-Tong and Rosenstock, Dario and Liu, Bao-Jie and Somoroff, Aaron and Manucharyan, Vladimir E. and Wang, Chen},
  journal = {Nature Communications},
  year    = {2026},
  volume  = {17},
  issue   = {1},
  pages   = {3209},
  doi     = {10.1038/s41467-026-69910-2},
  url     = {https://doi.org/10.1038/s41467-026-69910-2}
}

@article{Larson2026,
  title   = {Localized quasiparticles in a fluxonium with quasi-two-dimensional amorphous kinetic inductors},
  author  = {Larson, Trevyn F. Q. and Jones, Sarah Garcia and Kalmár, Tamás and Sanchez, Pablo Aramburu and Chitta, Sai Pavan and Verma, Varun and Genter, Kristen L. and Gill, Stephen T. and Cicak, Katarina and Nam, Sae Woo and others},
  journal = {Nature Communications},
  year    = {2026},
  volume  = {17},
  issue   = {1},
  pages   = {3022},
  doi     = {10.1038/s41467-026-69709-1},
  url     = {https://doi.org/10.1038/s41467-026-69709-1}
}

@article{Connolly2024,
  title   = {Coexistence of Nonequilibrium Density and Equilibrium Energy Distribution of Quasiparticles in a Superconducting Qubit},
  author  = {Connolly, Thomas and Kurilovich, Pavel D. and Diamond, Spencer and Nho, Heekun and B\o{}ttcher, Charlotte G. L. and Glazman, Leonid I. and Fatemi, Valla and Devoret, Michel H.},
  journal = {Phys. Rev. Lett.},
  year    = {2024},
  volume  = {132},
  issue   = {21},
  pages   = {217001},
  doi     = {10.1103/PhysRevLett.132.217001},
  url     = {https://link.aps.org/doi/10.1103/PhysRevLett.132.217001}
}

@article{Marchegiani2025,
  title = {Nonequilibrium regimes for quasiparticles in superconducting qubits with gap-asymmetric junctions},
  author = {Marchegiani, Giampiero and Catelani, Gianluigi},
  journal = {Communications Physics},
  year = {2025},
  volume = {8},
  issue = {1},
  pages = {120},
  doi = {10.1038/s42005-025-02052-x},
  url = {https://doi.org/10.1038/s42005-025-02052-x}
}

@article{Dolan1977,
  title   = {Offset masks for lift-off photoprocessing},
  author  = {Dolan, G. J.},
  journal = {Appl. Phys. Lett.},
  year    = {1977},
  volume  = {31},
  issue   = {5},
  pages   = {337--339},
  doi     = {10.1063/1.89690},
  url     = {https://doi.org/10.1063/1.89690}
}

@article{Lecocq2011,
  title   = {Junction fabrication by shadow evaporation without a suspended bridge},
  author  = {Lecocq, Florent and Pop, Ioan M. and Peng, Zhihui and Matei, Iulian and Crozes, Thierry and Fournier, Thierry and Naud, Cécile and Guichard, Wiebke and Buisson, Olivier},
  journal = {Nanotechnology},
  year    = {2011},
  volume  = {22},
  issue   = {31},
  pages   = {315302},
  doi     = {10.1088/0957-4484/22/31/315302},
  url     = {https://doi.org/10.1088/0957-4484/22/31/315302}
}

@article{Glazman2021,
  title = {Bogoliubov quasiparticles in superconducting qubits},
  author = {Glazman, Leonid and Catelani, Gianluigi},
  journal = {SciPost Phys. Lect. Notes},
  year = {2021},
  pages = {31},
  doi = {10.21468/SciPostPhysLectNotes.31},
  url = {https://scipost.org/SciPostPhysLectNotes.31}
}

@article{Yelton2024,
  title   = {Modeling phonon-mediated quasiparticle poisoning in superconducting qubit arrays},
  author  = {Yelton, E. and Larson, C. P. and Iaia, V. and Dodge, K. and La Magna, G. and Baity, P. G. and Pechenezhskiy, I. V. and McDermott, R. and Kurinsky, N. A. and Catelani, G. and others},
  journal = {Phys. Rev. B},
  year    = {2024},
  volume  = {110},
  issue   = {2},
  pages   = {024519},
  doi     = {10.1103/PhysRevB.110.024519},
  url     = {https://link.aps.org/doi/10.1103/PhysRevB.110.024519}
}

@article{Watanabe2025,
  title   = {Nondemolition fluorescence readout and high-fidelity unconditional reset of a fluxonium qubit via dissipation engineering},
  author  = {Watanabe, Shu and Hida, Kotaro and Matsuura, Kohei and Nakamura, Yasunobu},
  journal = {Phys. Rev. A},
  year    = {2025},
  volume  = {112},
  issue   = {1},
  pages   = {012624},
  doi     = {10.1103/gcgj-srgy},
  url     = {https://link.aps.org/doi/10.1103/gcgj-srgy}
}

@article{Gustavsson2016,
  title   = {Suppressing relaxation in superconducting qubits by quasiparticle pumping},
  author  = {Gustavsson, Simon and Yan, Fei and Catelani, Gianluigi and Bylander, Jonas and Kamal, Archana and Birenbaum, Jeffrey and Hover, David and Rosenberg, Danna and Samach, Gabriel and Sears, Adam P. and others},
  journal = {Science},
  year    = {2016},
  volume  = {354},
  issue   = {6319},
  pages   = {1573--1577},
  doi     = {10.1126/science.aah5844},
  url     = {https://doi.org/10.1126/science.aah5844}
}

@article{Spiecker2023,
  title   = {Two-level system hyperpolarization using a quantum {Szilard} engine},
  author  = {Spiecker, Martin and Paluch, Patrick and Gosling, Nicolas and Drucker, Niv and Matityahu, Shlomi and Gusenkova, Daria and G{\"u}nzler, Simon and Rieger, Dennis and Takmakov, Ivan and Valenti, Francesco and others},
  journal = {Nature Physics},
  year    = {2023},
  volume  = {19},
  issue   = {9},
  pages   = {1320--1325},
  doi     = {10.1038/s41567-023-02082-8},
  url     = {https://doi.org/10.1038/s41567-023-02082-8}
}

@article{Lisenfeld2016,
  title   = {Decoherence spectroscopy with individual two-level tunneling  defects},
  author  = {Lisenfeld, J{\"u}rgen and Bilmes, Alexander and Matityahu, Shlomi and Zanker, Sebastian and Marthaler, Michael and Schechter, Moshe and Sch{\"o}n, Gerd and Shnirman, Alexander and Weiss, Georg and Ustinov, Alexey V.},
  journal = {Scientific Reports},
  year    = {2016},
  volume  = {6},
  issue   = {1},
  pages   = {23786},
  doi     = {10.1038/srep23786},
  url     = {https://doi.org/10.1038/srep23786}
}

@article{Bilmes2022,
  title   = {Probing defect densities at the edges and inside {Josephson} junctions of superconducting qubits},
  author  = {Bilmes, Alexander and Volosheniuk, Serhii and Ustinov, Alexey V. and Lisenfeld, J{\"u}rgen},
  journal = {npj Quantum Information},
  year    = {2022},
  volume  = {8},
  issue   = {1},
  pages   = {24},
  doi     = {10.1038/s41534-022-00532-4},
  url     = {https://doi.org/10.1038/s41534-022-00532-4}
}

@article{Sun2023,
  title   = {Characterization of Loss Mechanisms in a Fluxonium Qubit},
  author  = {Sun, Hantao and Wu, Feng and Ku, Hsiang-Sheng and Ma, Xizheng and Qin, Jin and Song, Zhijun and Wang, Tenghui and Zhang, Gengyan and Zhou, Jingwei and Shi, Yaoyun and others},
  journal = {Phys. Rev. Appl.},
  year    = {2023},
  volume  = {20},
  issue   = {3},
  pages   = {034016},
  doi     = {10.1103/PhysRevApplied.20.034016},
  url     = {https://link.aps.org/doi/10.1103/PhysRevApplied.20.034016}
}

@article{Houzet2019,
  title = {Photon-Assisted Charge-Parity Jumps in a Superconducting Qubit},
  author = {Houzet, M. and Serniak, K. and Catelani, G. and Devoret, M. H. and Glazman, L. I.},
  journal = {Phys. Rev. Lett.},
  year = {2019},
  volume = {123},
  issue = {10},
  pages = {107704},
  doi = {10.1103/PhysRevLett.123.107704},
  url = {https://link.aps.org/doi/10.1103/PhysRevLett.123.107704}
}

@article{Diamond2022,
  title = {Distinguishing Parity-Switching Mechanisms in a Superconducting Qubit},
  author = {Diamond, S. and Fatemi, V. and Hays, M. and Nho, H. and Kurilovich, P. D. and Connolly, T. and Joshi, V. R. and Serniak, K. and Frunzio, L. and Glazman, L. I. and others},
  journal = {PRX Quantum},
  year = {2022},
  volume = {3},
  issue = {4},
  pages = {040304},
  doi = {10.1103/PRXQuantum.3.040304},
  url = {https://link.aps.org/doi/10.1103/PRXQuantum.3.040304}
}

@article{Zhu2013,
  title   = {Circuit {QED} with fluxonium qubits: Theory of the dispersive regime},
  author  = {Zhu, Guanyu and Ferguson, David G. and Manucharyan, Vladimir E. and Koch, Jens},
  journal = {Phys. Rev. B},
  year    = {2013},
  volume  = {87},
  issue   = {2},
  pages   = {024510},
  doi     = {10.1103/PhysRevB.87.024510},
  url     = {https://link.aps.org/doi/10.1103/PhysRevB.87.024510}
}

@misc{Yang2026,
  title         = {High-Temporal-Resolution Measurements of the Impacts of Ionizing Radiation on Superconducting Qubits},
  author        = {Yang, Jihee and Carroll, Thomas J. and Mason, Philip and Schwartz, Robert and O'Hara, Kenneth M. and Lund, Jennifer and Gottschalk, Michael and Stephenson, Timothy and Friedman, Lawrence H. and Yumiceva, Francisco and others},
  year          = {2026},
  eprint        = {2602.23544},
  archivePrefix = {arXiv}
}

@article{Wang2014,
  title   = {Measurement and control of quasiparticle dynamics in a superconducting qubit},
  author  = {Wang, C. and Gao, Y. Y. and Pop, I. M. and Vool, U. and Axline, C. and Brecht, T. and Heeres, R. W. and Frunzio, L. and Devoret, M. H. and Catelani, G. and others},
  journal = {Nature Communications},
  year    = {2014},
  volume  = {5},
  issue   = {1},
  pages   = {5836},
  doi     = {10.1038/ncomms6836},
  url     = {https://doi.org/10.1038/ncomms6836}
}

@article{Guruswamy2015,
  title = {Nonequilibrium Superconducting Thin Films with Sub-gap and Pair-breaking Photon Illumination},
  author = {Guruswamy, T. and Goldie, D. J. and Withington, S.},
  journal = {Superconductor Science and Technology},
  year = {2015},
  volume = {28},
  issue  = {5},
  pages = {054002},
  doi = {10.1088/0953-2048/28/5/054002},
  url = {https://doi.org/10.1088/0953-2048/28/5/054002}
}

@article{Fischer2024,
  title = {Nonequilibrium Quasiparticle Distribution in Superconducting Resonators: Effect of Pair-Breaking Photons},
  author = {Fischer, Paul B. and Catelani, Gianluigi},
  journal = {SciPost Physics},
  year = {2024},
  volume = {17},
  issue  = {3},
  pages = {070},
  doi = {10.21468/SciPostPhys.17.3.070},
  url = {https://www.scipost.org/SciPostPhys.17.3.070}
}

@article{Antonenko2026,
  title   = {Effect of quasiparticles on the parameters of a gap-engineered transmon},
  author  = {Antonenko, Daniil S. and Kurilovich, Pavel D. and Matute-Ca\~nadas, Francisco J. and Glazman, Leonid I.},
  journal = {Phys. Rev. B},
  year    = {2026},
  volume  = {113},
  issue   = {5},
  pages   = {054504},
  doi     = {10.1103/t448-147x},
  url     = {https://link.aps.org/doi/10.1103/t448-147x}
}

@misc{Azar2026Th,
  title         = {Numerical Modeling of Quasiparticle-Induced Dissipation in Fluxonium Qubits},
  author        = {Azar, Kate and Hays, Max and Serniak, Kyle},
  year          = {2026},
  eprint        = {2607.24946},
  archivePrefix = {arXiv}
}

@article{Bista2026,
  title = {Readout-induced leakage of the fluxonium qubit},
  author = {Bista, Aayam and Thibodeau, Matthew and Nie, Ke and Chow, Kaicheung and Clark, Bryan K. and Kou, Angela},
  journal = {Phys. Rev. Appl.},
  year = {2026},
  volume = {25},
  issue = {3},
  pages = {034058},
  doi = {10.1103/wjdb-4814},
  url = {https://link.aps.org/doi/10.1103/wjdb-4814}
}

@misc{Chapple2026,
  title         = {Measurement-induced state transitions across the fluxonium qubit landscape},
  author        = {Chapple, Alex A. and Varbanov, Boris M. and McDonald, Alexander and Blais, Alexandre},
  year          = {2026},
  eprint        = {2604.08515},
  archivePrefix = {arXiv}
}

@misc{Zwanenburg2026,
  title         = {Experimental Characterization and Modeling of Measurement-Induced State-Transitions in a Fluxonium Superconducting Qubit},
  author        = {Zwanenburg, Martijn F. S. and Hu, Jinlun and Huang, Eugene Y. and Yilmaz, Figen and Singh, Siddharth and Andersen, Christian Kraglund},
  year          = {2026},
  eprint        = {2606.17866},
  archivePrefix = {arXiv}
}

@article{Liu2024,
  title   = {Quasiparticle Poisoning of Superconducting Qubits from Resonant Absorption of Pair-Breaking Photons},
  author  = {Liu, C. H. and Harrison, D. C. and Patel, S. and Wilen, C. D. and Rafferty, O. and Shearrow, A. and Ballard, A. and Iaia, V. and Ku, J. and Plourde, B. L. T. and others},
  journal = {Phys. Rev. Lett.},
  year    = {2024},
  volume  = {132},
  issue   = {1},
  pages   = {017001},
  doi     = {10.1103/PhysRevLett.132.017001},
  url     = {https://link.aps.org/doi/10.1103/PhysRevLett.132.017001}
}

@article{Nho2026,
  title   = {Recovery Dynamics of a Gap-Engineered Transmon after a Quasiparticle Burst},
  author  = {Nho, Heekun and Connolly, Thomas and Kurilovich, Pavel D. and Diamond, Spencer and B{\o}ttcher, Charlotte G. L. and Glazman, Leonid I. and Devoret, Michel H.},
  journal = {Phys. Rev. Lett.},
  year    = {2026},
  volume  = {136},
  issue   = {5},
  pages   = {050601},
  doi     = {10.1103/ql6q-wfpn},
  url     = {https://link.aps.org/doi/10.1103/ql6q-wfpn}
}

@article{Cassidy2017,
  title = {Demonstration of an ac {Josephson} junction laser},
  author = {Cassidy, M. C. and Bruno, A. and Rubbert, S. and Irfan, M. and Kammhuber, J. and Schouten, R. N. and Akhmerov, A. R. and Kouwenhoven, L. P.},
  journal = {Science},
  year = {2017},
  volume = {355},
  issue = {6328},
  pages = {939--942},
  doi = {10.1126/science.aah6640},
  url = {https://www.science.org/doi/abs/10.1126/science.aah6640}
}

@misc{Rafferty2021,
  title = {Spurious antenna modes of the transmon qubit},
  author = {Rafferty, O. and Patel, S. and Liu, C. H. and Abdullah, S. and Wilen, C. D. and Harrison, D. C. and McDermott, R.},
  year = {2021},
  eprint = {2103.06803},
  archivePrefix = {arXiv}
}

@article{Pan2022,
  title = {Engineering superconducting qubits to reduce quasiparticles and charge noise},
  author = {Pan, Xianchuang and Zhou, Yuxuan and Yuan, Haolan and Nie, Lifu and Wei, Weiwei and Zhang, Libo and Li, Jian and Liu, Song and Jiang, Zhi Hao and Catelani, Gianluigi and others},
  journal = {Nature Communications},
  year = {2022},
  volume = {13},
  issue  = {1},
  pages = {7196},
  doi = {10.1038/s41467-022-34727-2},
  url = {https://doi.org/10.1038/s41467-022-34727-2}
}

@article{Holst1994,
  title = {Effect of a Transmission Line Resonator on a Small Capacitance Tunnel Junction},
  author = {Holst, T. and Esteve, D. and Urbina, C. and Devoret, M. H.},
  journal = {Phys. Rev. Lett.},
  year = {1994},
  volume = {73},
  issue = {25},
  pages = {3455--3458},
  doi = {10.1103/PhysRevLett.73.3455},
  url = {https://link.aps.org/doi/10.1103/PhysRevLett.73.3455}
}

@misc{Lin2026,
  title = {Suppression of Quasiparticle Poisoning to $10^{-11}$ Levels in Superconducting Qubits via Infrared Shielding},
  author = {Lin, Wei-En and Ma, Chen-Hsun and Yeh, Erh-Hsiang and Peng, Wei-Lun and Wei, Yu-Sen and Goan, Hsi-Sheng and Wu, Cen-Shawn and Ke, Chung-Ting and Chen, Yung-Fu and Chen, Chii-Dong},
  year = {2026},
  eprint = {2606.07339},
  archivePrefix = {arXiv}
}

@article{Nguyen2022,
  title = {Blueprint for a High-Performance Fluxonium Quantum Processor},
  author = {Nguyen, Long B. and Koolstra, Gerwin and Kim, Yosep and Morvan, Alexis and Chistolini, Trevor and Singh, Shraddha and Nesterov, Konstantin N. and J\"unger, Christian and Chen, Larry and Pedramrazi, Zahra and others},
  journal = {PRX Quantum},
  year = {2022},
  volume = {3},
  issue = {3},
  pages = {037001},
  doi = {10.1103/PRXQuantum.3.037001},
  url = {https://link.aps.org/doi/10.1103/PRXQuantum.3.037001}
}

@article{Koch2007,
  title = {Charge-insensitive qubit design derived from the {Cooper} pair box},
  author = {Koch, Jens and Yu, Terri M. and Gambetta, Jay and Houck, A. A. and Schuster, D. I. and Majer, J. and Blais, Alexandre and Devoret, M. H. and Girvin, S. M. and Schoelkopf, R. J.},
  journal = {Phys. Rev. A},
  year = {2007},
  volume = {76},
  issue = {4},
  pages = {042319},
  doi = {10.1103/PhysRevA.76.042319},
  url = {https://link.aps.org/doi/10.1103/PhysRevA.76.042319}
}

@article{Krause2024,
  title = {Quasiparticle effects in magnetic-field-resilient three-dimensional transmons},
  author = {Krause, J. and Marchegiani, G. and Janssen, L. M. and Catelani, G. and Ando, Yoichi and Dickel, C.},
  journal = {Phys. Rev. Appl.},
  year = {2024},
  volume = {22},
  issue = {4},
  pages = {044063},
  doi = {10.1103/PhysRevApplied.22.044063},
  url = {https://link.aps.org/doi/10.1103/PhysRevApplied.22.044063}
}

@article{Randeria2024,
  title = {Dephasing in Fluxonium Qubits from Coherent Quantum Phase Slips},
  author = {Randeria, Mallika T. and Hazard, Thomas M. and Di Paolo, Agustin and Azar, Kate and Hays, Max and Ding, Leon and An, Junyoung and Gingras, Michael and Niedzielski, Bethany M. and Stickler, Hannah and others},
  journal = {PRX Quantum},
  year = {2024},
  volume = {5},
  issue = {3},
  pages = {030341},
  doi = {10.1103/PRXQuantum.5.030341},
  url = {https://link.aps.org/doi/10.1103/PRXQuantum.5.030341}
}

@article{Singh2025,
  title = {Impact of {Josephson}-Junction Array Modes on Fluxonium Readout},
  author = {Singh, Shraddha and Refael, Gil and Clerk, Aashish and Rosenfeld, Emma},
  journal = {PRX Quantum},
  year = {2025},
  volume = {6},
  issue = {4},
  pages = {040304},
  doi = {10.1103/brdj-ggfj},
  url = {https://link.aps.org/doi/10.1103/brdj-ggfj}
}

@article{Ardati2024,
  title = {Using Bifluxon Tunneling to Protect the Fluxonium Qubit},
  author = {Ardati, Wa\"el and L\'eger, S\'ebastien and Kumar, Shelender and Suresh, Vishnu Narayanan and Nicolas, Dorian and Mori, Cyril and D'Esposito, Francesca and Vakhtel, Tereza and Buisson, Olivier and Ficheux, Quentin and others},
  journal = {Phys. Rev. X},
  year = {2024},
  volume = {14},
  issue = {4},
  pages = {041014},
  doi = {10.1103/PhysRevX.14.041014},
  url = {https://link.aps.org/doi/10.1103/PhysRevX.14.041014}
}

@misc{Marchegiani2025arXiv,
  title = {Proximity effect in asymmetric-gap superconducting bilayers and regularization of transition rates},
  author = {Marchegiani, G. and Catelani, G.},
  year = {2025},
  eprint = {2512.17765},
  archivePrefix = {arXiv}
}

\end{document}